\documentclass[acmtog]{acmart}

\usepackage{booktabs} 
\usepackage{multirow}
\usepackage{makecell}
\usepackage{amsmath}

\makeatletter
\@namedef{ver@everyshi.sty}{}
\makeatother

\usepackage{colortbl}
\usepackage{graphicx}
\usepackage{subcaption}
\usepackage{lipsum}
\usepackage{arydshln}

\newcommand{\best}{\cellcolor{tablered}}
\newcommand{\sbest}{\cellcolor{orange}}
\newcommand{\tbest}{\cellcolor{yellow}}

\newcommand{\bn}{\mathbf{n}}

\makeatletter
\DeclareRobustCommand\onedot{\futurelet\@let@token\@onedot}
\def\@onedot{\ifx\@let@token.\else.\null\fi\xspace}

\makeatother

\definecolor{yellow}{rgb}{1, 1, 0.7}
\definecolor{orange}{rgb}{1, 0.85, 0.7}
\definecolor{tablered}{rgb}{1, 0.7, 0.7}
\definecolor{red}{rgb}{1, 0, 0}

\definecolor{wincolor}{rgb}{0.85, 0.0, 0.0}

\definecolor{darkyellow}{rgb}{0.8, 0.8, 0.5}
\definecolor{darkred}{rgb}{0.7, 0.3, 0.3}
\definecolor{darkgreen}{rgb}{0.3, 0.7, 0.3}
\definecolor{green}{rgb}{0, 1.0, 0}
\definecolor{pink}{rgb}{1, 0.4, 0.7}

\definecolor{realred}{rgb}{0.95, 0.1, 0.0}

\usepackage{xcolor}

\usepackage{algpseudocode}
\usepackage{wrapfig}
\usepackage[ruled]{algorithm2e} 

\SetAlFnt{\small}
\SetAlCapFnt{\small}
\SetAlCapNameFnt{\small}
\SetAlCapHSkip{0pt}

\acmJournal{TOG}

\definecolor{myPurple}{rgb}{0.4, .0, .8}
\definecolor{myPurple2}{rgb}{0.4, .0, .5}
\definecolor{myGreen}{rgb}{0, .8, .3}
\definecolor{myRed}{rgb}{0.8, .2, .2}
\definecolor{myOrange}{rgb}{0.7, 0.45, 0.2}
\definecolor{myBlue}{rgb}{.0, .0, 1.0}
\definecolor{myBlue2}{rgb}{.0, .0, 0.5}
\definecolor{myBlack}{rgb}{.0, .0, 0.0}

\begin{document}
\title{RaDe-GS: Rasterizing Depth in Gaussian Splatting}

\author{Baowen Zhang}
\affiliation{%
 \institution{Hong Kong University of Science and Technology}
 \city{Hong Kong}
 \country{Hong Kong}
 }
\email{bzhangcm@connect.ust.hk}
\author{Chuan Fang}
\affiliation{%
 \institution{Hong Kong University of Science and Technology}
 \city{Hong Kong}
 \country{Hong Kong}
}
\email{cfangac@connect.ust.hk}
\author{Rakesh Shrestha}
\affiliation{%
\institution{Simon Fraser University}
\city{Burnaby}
\country{Canada}
}
\email{rakeshs@sfu.ca}
\author{Yixun Liang}
\affiliation{%
 \institution{Hong Kong University of Science and Technology}
 \city{Hong Kong}
 \country{Hong Kong}
 }
\email{lyxun2000@gmail.com}
\author{Xiaoxiao Long}
\affiliation{%
 \institution{Hong Kong University of Science and Technology}
 \city{Hong Kong}
 \country{Hong Kong}
}
\email{xxlong@connect.hku.hk}
\author{Ping Tan}
\affiliation{%
 \institution{Hong Kong University of Science and Technology}
 \city{Hong Kong}
 \country{Hong Kong}
}
\affiliation{%
\institution{Simon Fraser University}
\city{Burnaby}
\country{Canada}
}
\email{pingtan@ust.hk}

\begin{abstract}
Gaussian Splatting (GS) has proven to be highly effective in novel view synthesis, achieving high-quality and real-time rendering. However, its potential for reconstructing detailed 3D shapes has not been fully explored. Existing methods often suffer from limited shape accuracy due to the discrete and unstructured nature of Gaussian splats, which complicates the shape extraction. 
While recent techniques like 2D GS have attempted to improve shape reconstruction, they often reformulate the Gaussian primitives in ways that reduce both rendering quality and computational efficiency. To address these problems, our work introduces a rasterized approach to render the depth maps and surface normal maps of general 3D Gaussian splats.
Our method not only significantly enhances shape reconstruction accuracy but also maintains the computational efficiency intrinsic to Gaussian Splatting. It achieves a Chamfer distance error comparable to NeuraLangelo\cite{li2023neuralangelo} on the DTU dataset and maintains similar computational efficiency as the original 3D GS methods.
Our method is a significant advancement in Gaussian Splatting and can be directly integrated into existing Gaussian Splatting-based methods.

\end{abstract}

%
%


\begin{teaserfigure}
\vspace{-0.3cm}
    {\large \textcolor{magenta}{\texttt{\href{https://baowenz.github.io/radegs/}{https://baowenz.github.io/radegs}}}}\\
  \vspace{0.2cm}
  \includegraphics[width=\textwidth]{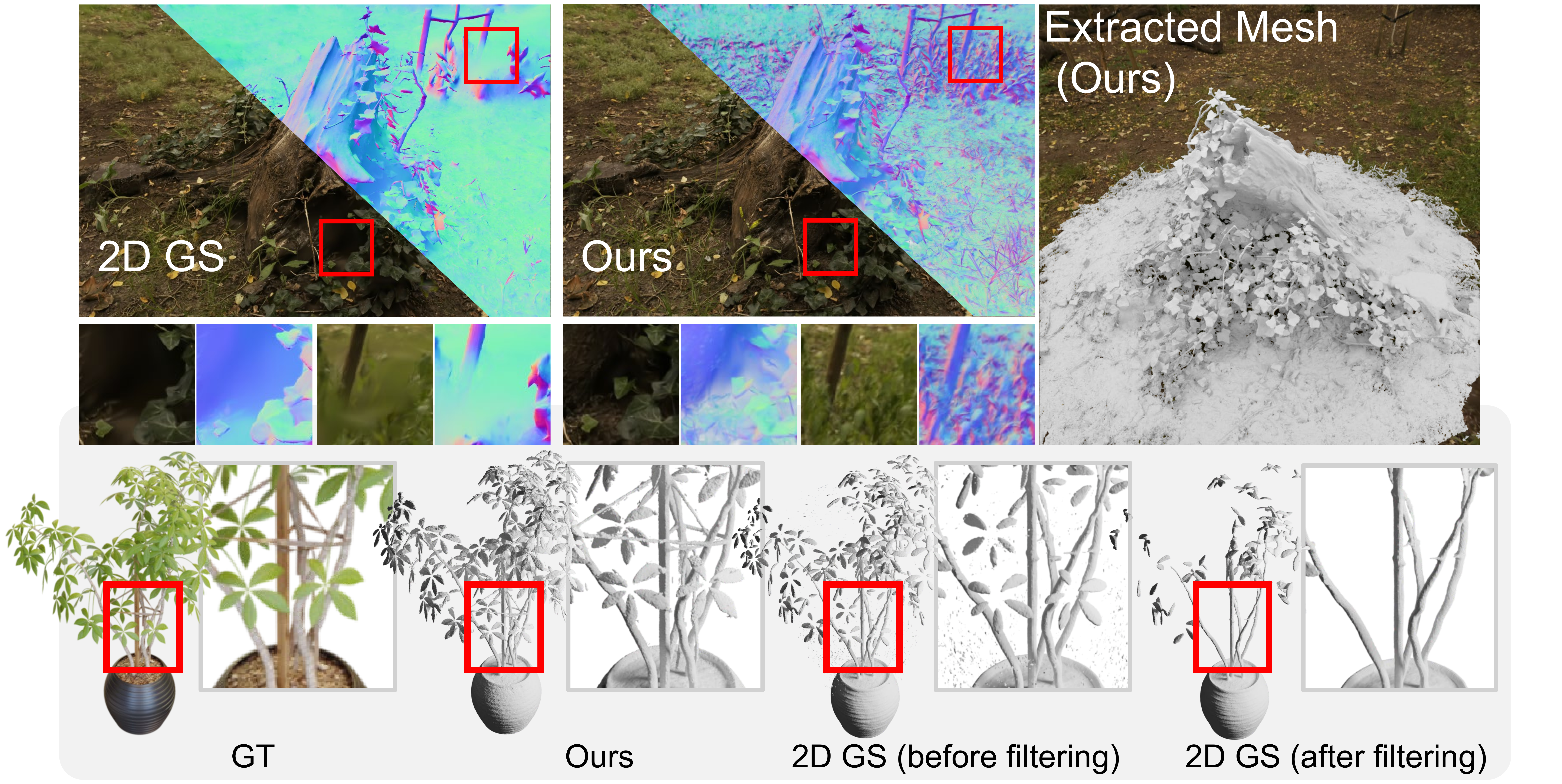}
  \vspace{-0.8cm}
  \caption{We present a rasterized method to compute the depth and surface normal maps of general Gaussian splats. Our method achieves high-quality 3D shape reconstruction and maintains excellent training and rendering efficiency. In comparison, forcing Gaussian splats to be planar as in 2D GS~\cite{huang20242d} produces blurry novel view rendering and noisy 3D shape. 2D GS  by default filters the noise after mesh extraction.}
  \label{fig:teaser}
\end{teaserfigure}

\maketitle

\begin{figure*}[t]
    \centering
    \setlength{\abovecaptionskip}{0.cm}
    \includegraphics[width=1\linewidth]{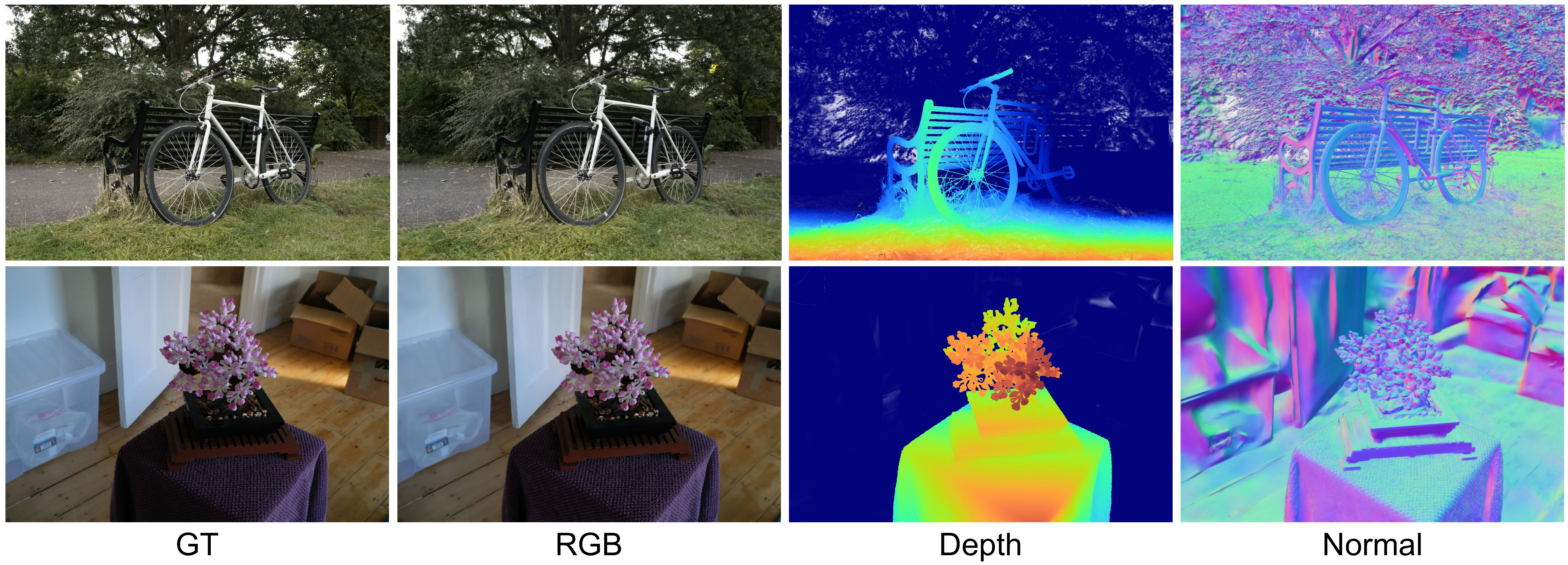}
    \caption{Quality results on Mip-NeRF 360 dataset. From left to right, the images are ground truth color images, rendered color images, rendered depth maps, and rendered normal maps, respectively.
    }
    \label{fig:ours_nerfsyn_nvs_recons}
    \vspace{-0.1in}
\end{figure*}
\section{Introduction}

3D reconstruction from multi-view images is a classic problem with numerous applications in computer vision and graphics. This task typically involves generating depth maps through the multi-view stereo algorithms that utilize sophisticated optimization methods \cite{Kolmogorov2001, Sun2003, Bleyer2011} or pretrained neural networks \cite{yao2018mvsnet,gu2019cas}. The depth maps estimated from different viewpoints can then be integrated \cite{newcombe2011kinectfusion} to create a complete triangle mesh model. Although this traditional image-based modeling approach \cite{fuhrmann2014mve} delivers accurate results, it has limited robustness, particularly on reflective and shiny surfaces.

Neural Radiance Field (NeRF) \cite{mildenhall2021nerf} employs an implicit representation of 3D scenes and achieves photorealist results for novel-view rendering through an analysis-by-synthesis approach. Despite its success, the original NeRF method tends to produce biased depth maps and noisy 3D reconstruction. NeuS \cite{wang2021neus} incorporates a Signed Distance Function (SDF) into the NeRF framework, significantly improving the accuracy of depth and shape reconstructions. These improvements in 3D reconstruction accuracy are further enhanced by hierarchical volumetric features in NeuraLangelo \cite{li2023neuralangelo}. However, the implicit representation of 3D scenes requires ray tracing to render images, which makes the training of NeRF computationally inefficient. Consequently, these methods \cite{mildenhall2021nerf, wang2021neus, li2023neuralangelo} often take several hours to optimize a 3D model from a set of input images. 

Gaussian Splatting (GS) \cite{kerbl20233d} introduces an explicit representation for more efficient optimization and rendering. It represents a 3D scene using a set of translucent Gaussian spheres, which can be rendered efficiently by rasterization and can reconstruct a 3D scene in minutes. However, this representation complicates the computation of SDFs, as analyzed in \cite{guedon2023sugar}. Consequently, it is hard to extract 3D surfaces from GSs, which are necessary for applications like simulation and obstacle detection. 
Some recent works \cite{huang20242d,guedon2023sugar} attempt to make the Gaussian spheres planar to facilitate surface extraction.
However, this lower dimensional 
representation leads to optimization challenges, especially for complicated shapes, as shown in the grasses and stump in Figure~\ref{fig:teaser}. 
In general, 2D GS methods \cite{huang20242d,guedon2023sugar} decrease the Peak Signal-to-Noise Ratio (PSNR), as evidenced in Table~\ref{tab:mipnerf360} and Table~\ref{tab:nerf}. 
To address these challenges, GOF \cite{yu2024gaussian} introduces a ray-tracing based method to compute the opacity along light rays to extract high-quality surfaces. While this approach generates excellent surface reconstruction, the ray-tracing process costs significant computational overhead. 
For example, GOF requires about one hour to optimize a scene from the DTU dataset \cite{jensen2014large}, while the standard GS method takes only about 5.2 minutes, as shown in Table~\ref{tab:dtu_result}.

We develop a rasterized method to compute depth maps for general Gaussian splats. Our method enjoys similar computation efficiency as the standard GS \cite{kerbl20233d} thanks to the rasterized approach in computing depth and normal maps. 
On the DTU dataset, our method achieves a shape reconstruction accuracy of 0.69 mm Chamfer Distance after 5-minute training. This performance matches that of the implicit method NeuraLangelo, which has an accuracy of 0.61 mm, and exceeds that of recent GS-based methods like GOF (0.74 mm) and 2D GS (0.80 mm). 

We have derived a closed-form solution for the intersections of light rays and Gaussian splats. Specifically, we evaluate the Gaussian values along each light ray from the camera center. The intersecting point on each ray is identified as the point where the Gaussian values are maximized. 
These intersection points between a Gaussian splat and a bundle of light rays lie on a general curved surface, which defines the projected depth of a Gaussian splat on the image plane. We discovered that these intersection points are co-planar under the approximate affine projection \cite{zwicker2002ewa}. As a result, the projected depth of each Gaussian splat can be computed efficiently by rasterization according to our derived planar equation. 
The final depth map is computed as the median depth among the projected Gaussian splats, taking into account their translucency. Similarly, the surface normal map is derived through rasterized computation. This approach allows our method to produce accurate depth and normal maps, while maintaining the rendering and optimization efficiency of Gaussian splitting. As shown in Figure.~\ref{fig:ours_nerfsyn_nvs_recons}, our method can render high quality color, depth, and normal maps from Gaussian splats.

In summary, the main contribution of this work is a novel rasterized method for computing depth and normal maps tailored to general Gaussian splats. Extensive experimental evaluation demonstrates that our method achieves high-quality 3D reconstruction, comparable to those of the recent implicit method NeuraLangelo\cite{li2023neuralangelo}. Additionaly, it maintains rendering and optimization efficiency on par with the original 3D Gaussian Splats \cite{kerbl20233d}.

\section{Related Work}

\subsection{Stereo and Multi-view Stereo}
Stereo and multi-view stereo depth estimation are classical problems in computer vision. Traditional methods employ sophisticated optimization techniques, such as belief-propagation~\cite{Sun2003}, graph-cut~\cite{kolmogorov2004energy}, and PatchMatch~\cite{Bleyer2011,barnes2009patchmatch} to establish pixel correspondences. While these methods have achieved accurate 3D reconstructions, they often struggle with textureless regions. 
More recent works~\cite{Zbontar2015,Luo2016,Kendall2017} utilize neural networks for end-to-end depth map estimation. MVSNet~\cite{yao2018mvsnet} integrates learning-based cost-volume filtering in the classical plane-sweeping stereo algorithm~\cite{Collins1996}. Furthermore, CasMVSNet~\cite{gu2020cascade} designs a cascade cost-volume filtering to improve depth accuracy and reduce GPU memory usage. TransMVSNet~\cite{Ding2022} applies the attention mechanism in Transformers to improve feature correlation.
Stereo and multi-view stereo methods recover a depth map for each input image, which can be fused into a complete 3D model with truncated signed distance functions (TSDF)~\cite{Curless1996,Zach2007,newcombe2011kinectfusion}.

\subsection{Nerual Radiance Field}
Unlike multi-view stereo algorithms that rely on depth map representation, NeRF~\cite{mildenhall2021nerf} parameterizes a scene with an MLP network that maps spatial coordinates to color and translucency values. 
NeRF achieves photo-realistic novel viewpoint rendering by optimizing network parameters through an analysis-by-synthesis approach. 
Subsequent developments have led to significant improvements. For anti-aliasing rendering, MipNeRF~\cite{barron2021mip} uses a conical frustum to represent the target scene at varying scales, and MipNeRF360~\cite{barron2022mip} further extends this method to unbounded scenes. Although these methods achieve faithful rendering, they require extensive training hours. Efforts to accelerate training include Plenoxels~\cite{fridovich2022plenoxels}, which uses a 3D grid of spherical harmonics, and Instant-NGP~\cite{muller2022instant}, which combines a hashed feature volume and a shallow MLP. Additional methods focus on improving rendering efficiency at the inference stage through techniques like baking~\cite{hedman2021baking} or distillation methods~\cite{yu2021plenoctrees,chen2023mobilenerf,reiser2023merf}.

Another important direction in NeRF research aims to improve the accuracy of reconstructed 3D shapes. Notable works\cite{wang2021neus,yariv2021volume,yu2022monosdf,wang2022neuris,li2023neuralangelo} have integrated a Signed Distance Function (SDF) with the radiance field to create high-fidelity 3D models.
NeuS \cite{wang2021neus} establishes the connection between translucency 
and SDF values to ensure unbiased surface reconstruction.
Moreover, Neuralangelo \cite{li2023neuralangelo} incorporates multi-resolution feature grids with SDF, achieving high-quality reconstruction on large-scale scenes.
Despite these improvements, NeRF-based methods typically require expensive optimization with substantial GPU hours and memory. For instance, Neuralangelo~\cite{li2023neuralangelo} requires about 128 GPU hours to reconstruct a single scene in the Tanks \& Temples dataset~\cite{Knapitsch2017}.

\subsection{Gaussian Splatting}

Gaussian Splatting \cite{kerbl20233d} employs a set of 3D Gaussian primitives to represent a 3D scene. Combined with rasterized rendering, it and many variant works~\cite{fan2023lightgaussian,lin2024vastgaussian,cheng2024gaussianpro,wang2024dc} achieve real-time rendering and fast optimization. 
Building on this success, Mip-Splatting \cite{yu2023mip} incorporates low-pass filters to address aliasing problems. LightGaussian \cite{fan2023lightgaussian} optimizes memory usages with a compact format for 3D Gaussians, while VastGaussian \cite{lin2024vastgaussian} scales Gaussian Splatting to larger scenes.

However, extracting 3D surfaces from Gaussian splats remains challenging due to their discrete and unstructured nature.
To overcome this challenge, SuGaR~\cite{guedon2023sugar} and NeuGS-~\cite{chen2023neusg} favor flat Gaussians that better align with object surfaces. GSDF~\cite{yu2024gsdf} proposes a dual-branch network combining the standard 3D GS with NeuS~\cite{wang2021neus} to improve rendering fidelity and reconstruction accuracy simultaneously. Moreover, post-processing and joint-optimization methods~\cite{guedon2023sugar} can improve the results, but at the cost of increased training time.
2D GS~\cite{huang20242d} directly replaces 3D Gaussian primitives with flat 2D Gaussians for effective surface reconstruction. Yet, it sacrifices novel-view synthesis capability and training stability because the 
2D Gaussians can lead to degenerate scene representation and fail to capture more complicated scenes.

Our work analyzes the depth evaluation in standard 3D Gaussian splats and proposes a novel rasterized method for depth map computation. Our method not only achieves precise surface reconstruction but also retains the rendering and optimization efficiency of 3D Gaussian splats.

\section{Method}

\subsection{Gaussain Splatting Preliminary}
The standard Gaussian Splatting~\cite{kerbl20233d} represents a 3D scene by a set of translucent 3D Gaussians. Each 3D Gaussian is defined as,
\begin{equation}
    G(\mathbf{x}) = e^{-(\mathbf{x}-\mathbf{x}_c)^\top \mathbf{\Sigma}^{-1}(\mathbf{x}-\mathbf{x}_c)},
\end{equation}
where $\mathbf{x}_c\in \mathbb{R}^3$ is the Gaussian center, and $\mathbf{\Sigma}\in \mathbb{R}^{3\times 3}$ is the covariance matrix. The covariance $\mathbf{\Sigma}$ is parameterized by a scaling matrix $\mathbf{S}$ and rotation matrix $\mathbf{R}$ as $\mathbf{\Sigma}=\mathbf{R}\mathbf{S}\mathbf{S}^\top \mathbf{R}^\top$.

\begin{figure}
    \centering
    \includegraphics[width=0.8\linewidth]{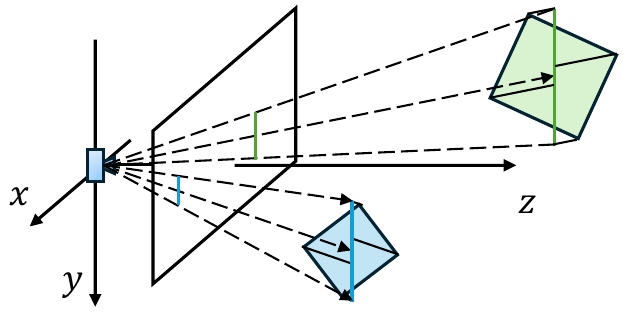}
    \caption{Illustration of the local affine projection in Gaussian Splatting~\cite{kerbl20233d,zwicker2002ewa}. It approximates the perspective projection by a local parallel projection at each Gaussian splat. 
    }
    \label{fig:intersections}
\end{figure}

\paragraph{Approximate Local Affine Projection}
Gaussian Splatting approximates the perspective camera projection locally by an affine transformation \cite{zwicker2002ewa} for each 3D Gaussian to achieve efficient rasterized rendering, as illustrated in Figure~\ref{fig:intersections}. The projected 2D Gaussian can be computed as,
\begin{equation}
    \mathbf{\Sigma}' = \mathbf{J}\mathbf{W}\mathbf{\Sigma}\mathbf{W}^\top\mathbf{J}^\top,
    \label{eq:ray_cov}
\end{equation}
where $\mathbf{\Sigma}'\in \mathbb{R}^{3\times 3}$ is the covariance matrix in camera coordinate frame, $\mathbf{W}$ is the rotation matrix from the world to camera coordinate system, and $\mathbf{J}$ is the Jacobian of the perspective transformation. The 2D Gaussian covariance is obtained by skipping the last row and column of $\Sigma'$.

\paragraph{Alpha Blending}
Gaussian Splatting sorts the projected 2D Gaussians by their depth and computes the color at each pixel by $\alpha$-blending,
\begin{equation}
    c = \sum_{i\in N} c_i\alpha_i \prod_{j=1}^{i-1}(1-\alpha_j),
\end{equation}
where $c$ is the rendered pixel color, $c_i$ is the color of the $i-$th Gaussian kernel computed from its spherical harmonics coefficients and viewing direction, $\alpha_i$ is the pixel translucency determined by the opacity of the $i-$th Gaussian kernel and the pixel's position.

\subsection{Rasterizing Depth for Gaussian Splats}

Standard Gaussian Splatting \cite{kerbl20233d} evaluates the depth of each 2D Gaussian by its center to sort them for alpha blending. This constant depth per Gaussian splats cannot capture shape details. Therefore, we aim to 
compute a spatially varying depth within the projected 2D Gaussian and derive a rasterized method for its efficient evaluation.

Assume $(u_c, v_c)$ is the center of the 2D Gaussian. For a pixel $(u,v)$ covered by the projected Gaussian, we compute its depth as,
\begin{equation}
    d = z_c + \mathbf{p} \begin{pmatrix}\Delta u\\
                                                \Delta v
                                                \end{pmatrix},
\label{eq:depth_render}
\end{equation}
where $z_c$ is the depth of the Gaussian center, $\Delta u = u_c - u$ and $\Delta v = v_c - v$ are the relative pixel positions. The $1\times 2$ vector $\mathbf{p}\in \mathbb{R}^2$ is determined by the Gaussian parameters and camera extrinsic parameters. This formulation enables rasterized computation of spatially varying depths within a projected Gaussian. We derive it in detail in this subsection.
 
\subsubsection{Depth Under Perspective Projection} 
To make the derivation easier to understand, we first introduce the concepts in the camera coordinates with perspective projection. As shown in Figure~\ref{fig:ray_plane} (a), consider a light ray leaving the camera center $\mathbf{o}$ with unit direction $\mathbf{v}$. A point on this ray is parameterized by the distance $t$ to $\mathbf{o}$ as,
\begin{equation}
    \mathbf{x}=\mathbf{o}+t\mathbf{v}.
\end{equation}
The Gaussian value on the ray can be computed as a function of $t$ as follows,
\begin{equation}
    G^1(t) = e^{-(\mathbf{o}+t\mathbf{v} -\mathbf{x}_c)^\top\mathbf{\Sigma}^{-1}(\mathbf{o}+t\mathbf{v}-\mathbf{x}_c)}.
    \label{eq:1DGaussian}
\end{equation}
According to Equation~\ref{eq:1DGaussian}, the Gaussian value along the ray is a 1D Gaussian function. 

We define the `intersection point' of the ray and the 3D Gaussian to be the point that maximizes the 1D Gaussian function  $G^1(t)$. As shown in Figure~\ref{fig:ray_plane} (a), the green point is the intersection of the ray with the Gaussian splat. The distance $t^*$ between the intersection point and the camera center can be computed in closed-form by locating the maximum value of $G^1(t)$ as,
\begin{equation}
    t^* = \frac{\mathbf{v}^\top \mathbf{\Sigma}^{-1}(\mathbf{x}_c -\mathbf{o})}{\mathbf{v}^\top \mathbf{\Sigma}^{-1} \mathbf{v}}.
    \label{eq:tstar}
\end{equation}

Equation~\ref{eq:tstar} implies that intersections of a 3D Gaussian and a bundle of light rays form a curved surface, where different pixels have different depth values of $t^*$ and different viewing direction $\mathbf{v}$.

\subsubsection{Depth Under Local Affine Projection} 
Now, we derive the pixel's depth under the local affine projection -- the projection model adopted in\cite{kerbl20233d,zwicker2002ewa}, where each 3D Gaussian undergoes an affine projection. We will show this simplified projection model allows for a rasterized method to compute the spatially varying depth of a projected Gaussian splat. 

\begin{figure}
    \centering
    \includegraphics[width=\linewidth]{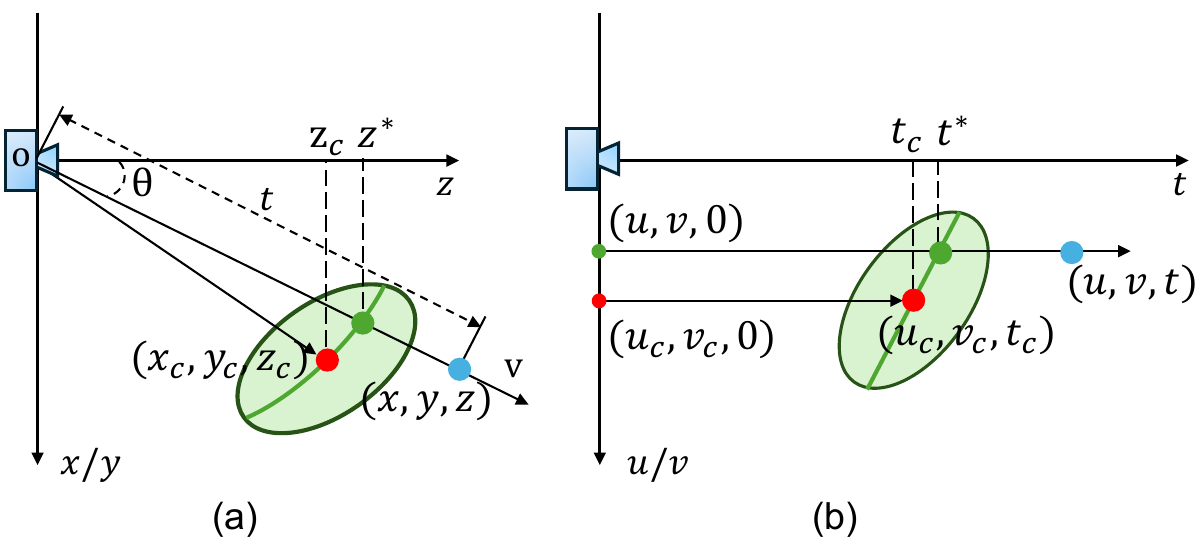}
    \caption{Intersecting a light ray with a 3D Gaussian in the `camera space' (a) and `ray space' (b). The Gaussian center $(x_c,y_c,z_c)$ is transformed to  $(u_c, v_c, t_c)$. A point $(x,y,z)$ is transformed to $(u,v,t)$. The green curve in (a) and the green line in (b) represent the set of intersections of the Gaussian with different light rays.}
    \label{fig:ray_plane}
\end{figure}

Under the local affine projection, each 3D Gaussian is undergoing an affine projection locally, as shown in Figure~\ref{fig:ray_plane} (b). Following \cite{zwicker2002ewa}, we refer to the coordinate system in Figure~\ref{fig:ray_plane} (b) as the `ray space', while the one in Figure~\ref{fig:ray_plane} (a) is the `camera space' for the clarity of discussion.

\noindent \textbf{Transformation from Camera to Ray Space.} 
Ray space is a non-Cartesian coordinate system that enables an easy formulation of our derivation.
The blue point in Figure~\ref{fig:ray_plane} (a), $\mathbf{x} = (x,y,z)^\top$ in the camera space, is transformed to the blue point in Figure~\ref{fig:ray_plane} (b), $\mathbf{u} = (u, v, t)$ in the ray space. The first two coordinates $(u, v)$ are the image plane coordinates, and $t$ is the distance between the point and the $uv$-plane. In other words, $t = \sqrt{x^2 + y^2 + z^2}$. Points on a light ray share the $u$ and $v$ coordinates but have varying distances $t$ to the camera center. Note the light ray direction $\mathbf{v}$ is normalized to $(0, 0, 1)^\top$ in the ray space.

The Gaussian Splats are transformed into the ray space, too. The transformed Gaussian function is,
\begin{equation}
    G'(\mathbf{u}) = e^{-(\mathbf{u}-\mathbf{u}_c)^\top \mathbf{\Sigma}'^{-1}(\mathbf{u}-\mathbf{u}_c)},
    \label{eq:cov_in_ray}
\end{equation}
where $\mathbf{u}$ is a point in ray space, $\mathbf{u}_c$  and $\mathbf{\Sigma}'$ are the transformed center and covariance matrix respectively. We denote the transformed Gaussian center as $\mathbf{u}_c = (u_c, v_c, t_c)^\top$. The transformed covariance matrix can be computed according to Equation~\ref{eq:ray_cov}.

\noindent \textbf{Intersection in Ray Space.} We derive the intersection in ray space by locating the maximum value of Gaussian on the light ray.
Similarly, in the ray space, a point is parameterized by its distance $t$ to the $uv$-plane as,
\begin{equation}
    \mathbf{u} = \mathbf{u}_o + t\mathbf{v}',
\label{eq:ray_in_ray}
\end{equation}
where $\mathbf{u}_o = (u, v, 0)^\top$ and $\mathbf{v}'=(0,0,1)^\top$. By substituting Equation~\ref{eq:ray_in_ray} into Equation~\ref{eq:cov_in_ray}, we can get the 1D Gaussian function defined on the light ray as,
\begin{equation}
    G'^1(t) = e^{-(\mathbf{\mathbf{u}_o+t\mathbf{v}'}-\mathbf{u}_c)^\top\mathbf{\Sigma}'^{-1}(\mathbf{\mathbf{u}_o+t\mathbf{v}'}-\mathbf{u}_c)}.
    \label{eq:1DGaussian_ray}
\end{equation}
Similarly, the maximum point can be located as,
\begin{equation}
    t^* = \frac{\mathbf{v}'^\top \mathbf{\Sigma}'^{-1}(\mathbf{u}_c -\mathbf{u}_o)}{\mathbf{v}'^\top \mathbf{\Sigma}'^{-1} \mathbf{v}'}.
    \label{eq:tstar_ray}
\end{equation}
While Equation~\ref{eq:tstar_ray} is similar to Equation~\ref{eq:tstar}, the direction $\mathbf{v}'$ here is a constant vector $(0,0,1)^\top$. Consequently, we can pre-compute $\mathbf{v}'^\top \mathbf{\Sigma}'^{-1} \mathbf{v}'$ and $\mathbf{v}'^\top \mathbf{\Sigma}'^{-1}$ for each Gaussian. In this way, the intersection point can be computed simply as
\begin{equation}
    t^* = \hat{\mathbf{q}} (\mathbf{u}_c -\mathbf{u}_o), 
    \label{eq:minus_form}
\end{equation}
where the $1\times 3$ vector $\hat{\mathbf{q}}$ is defined as,
\begin{equation}
    \hat{\mathbf{q}} = \frac{\mathbf{v}'^\top \mathbf{\Sigma}'^{-1}}{\mathbf{v}'^\top \mathbf{\Sigma}'^{-1} \mathbf{v}'}.
    \label{eq:q_hat}
\end{equation}
\noindent \textbf{Depth of Intersection.} We now derive the depth value of the intersection point, i.e. the green point in Figure~\ref{fig:ray_plane} (a) and (b). As shown in Figure~\ref{fig:ray_plane} (a), $t$ is the distance between the 3D point $\mathbf{x}$ and the camera center $\mathbf{o}$. So, the depth of $\mathbf{x}$ is simply, 
\begin{equation}
    d = cos\theta \ t^*,
\end{equation}
where $\theta$ is the angle between the light ray and the camera's principal axis. To simplify computation, we approximate it by $\theta_c$, the angle defined by the Gaussian center $\mathbf{x}_c$. Therefore, the depth of $\mathbf{x}$ becomes,
\begin{equation}
   d = cos \theta_c t^* = \frac{z_c}{t_c} t^* = \frac{z_c}{t_c}\hat{\mathbf{q}} (\mathbf{u}_c -\mathbf{u}_o) = \hat{\mathbf{p}} (\mathbf{u}_c -\mathbf{u}_o),
\label{eq:depth_unsimple}
\end{equation}
where $\hat{\mathbf{p}}=\frac{z_c}{t_c}\hat{\mathbf{q}}$ is a constant $1\times 3$ vector for a fixed Gaussian splat. 

We can further reformulate Equation~\ref{eq:depth_unsimple} as follows,
\begin{eqnarray}
    d = \hat{\mathbf{p}} (\mathbf{u}_c -\mathbf{u}_o) = 
    \hat{\mathbf{p}} \left(\begin{matrix}u_c - u\\v_c - v \\t_c\end{matrix} \right)
     = \hat{\mathbf{p}} \left(\begin{matrix}0 \\0 \\t_c\end{matrix}\right) + \hat{\mathbf{p}} \left(\begin{matrix}\Delta u \\\Delta v \\0\end{matrix}\right).   
     \label{eq:main_sub_depth}
\end{eqnarray}
We will simplify the first part $\hat{\mathbf{p}}(0,0,t_c)^T$ of Equation~\ref{eq:main_sub_depth}.
By replacing $\hat{\mathbf{p}}$ with $\frac{z_c}{t_c}\hat{\mathbf{q}}$ and substituting Equation~\ref{eq:q_hat} into it, the first part of Equation~\ref{eq:main_sub_depth} becomes:
\begin{equation}
\begin{split}
    \hat{\mathbf{p}} \left(\begin{matrix}0 \\0 \\t_c\end{matrix}\right)&=\frac{z_c}{t_c}\hat{\mathbf{q}}\left(\begin{matrix}0 \\0 \\t_c\end{matrix}\right)\\
    &=\frac{z_c}{t_c}\frac{\mathbf{v}'^\top \mathbf{\Sigma}'^{-1}}{\mathbf{v}'^\top \mathbf{\Sigma}'^{-1} \mathbf{v}'}\left(\begin{matrix}0 \\0 \\t_c\end{matrix}\right)\\
    &=\frac{z_c}{t_c}\ \frac{\mathbf{v}'^\top \mathbf{\Sigma}'^{-1}}{\mathbf{v}'^\top \mathbf{\Sigma}'^{-1} \mathbf{v}'}(t_c\mathbf{v}')\\
    &=\frac{z_c}{t_c}\ \frac{\mathbf{v}'^\top \mathbf{\Sigma}'^{-1}\mathbf{v}'}{\mathbf{v}'^\top \mathbf{\Sigma}'^{-1} \mathbf{v}'}t_c\\ &= z_c.
\end{split}
\end{equation}

For the second part $\hat{\mathbf{p}}(\Delta u,\Delta v, 0)^T$, we can skip the third element of $\hat{\mathbf{p}}$ to get $\mathbf{p}$ in Equation~\ref{eq:depth_render}. Thus, we obtain a rasterized method to compute the spatial varying depth for each pixel covered by the Gaussian splat.

\subsection{Rasterizing Normal for Gaussian Splats}
This section explains the computation of surface normal directions projected by a Gaussian splat. 
We first show that the intersection points form a plane in ray space as indicated by the green line segment in the Gaussian splat in Figure.~\ref{fig:ray_plane} (b). Therefore, we take the plane's normal direction as that of the projected Gaussian. We then transform the normal vector from the `ray space' to the `camera space' to compute the normal map.

\noindent \textbf{Plane Normal in Ray Space.} We derive the plane equation in ray space to get its normal direction. As we shown in the appendix, the intersection point in the ray space is,
\begin{equation}
\begin{split}
    \mathbf{u} &= 
    \left(\begin{array}{c}
    u \\
    v \\
    t^*
    \end{array}\right)=\left(\begin{array}{c}
    u \\
    v \\
    \frac{t_c}{z_c}d
    \end{array}\right)\\&=\left(\begin{array}{c}
    u_c - \Delta u \\
    v_c - \Delta v \\
    t_c +  (\Delta u \quad \Delta v)     \mathbf{q}^\top
    \end{array}\right)
    = \left(\begin{array}{c}
    u_c \\
    v_c \\
    t_c
    \end{array}\right) + 
    \left(\begin{array}{c}
    -\Delta u \\
    -\Delta v \\
     (\Delta u \quad \Delta v)     \mathbf{q}^\top
    \end{array}\right),
\end{split}
\label{eq:intsercting_point}
\end{equation}
where $\mathbf{q}=\frac{t_c}{z_c}\mathbf{p}$. 
Since $\hat{\mathbf{q}}=\frac{t_c}{z_c}\hat{\mathbf{p}}$ (Equation~15) and $\mathbf{p}$ contains the first two components of $\hat{\mathbf{p}}$, we know that $\mathbf{q}$ is the same as $\hat{\mathbf{q}}$ skipping the last component.
Note that $(u_c, v_c, t_c)^\top$ is the Gaussian center $\mathbf{u}_c$ and is a constant vector. Therefore, all the intersection points between the Gaussian splat and a bundle of light rays should lie on a plane in the ray space. The plane equation can be derived from Equation~\ref{eq:intsercting_point} as,
\begin{eqnarray}
 \left(\mathbf{u} - \mathbf{u}_c\right) 
 & = & \left(\begin{array}{c}
    -\Delta u \\
    -\Delta v \\
        (\Delta u \quad \Delta v)     \mathbf{q}^\top 
    \end{array}\right),
\\
\begin{pmatrix}
\mathbf{q} & 1
\end{pmatrix} \left( \mathbf{u} - \mathbf{u}_c \right) 
     &=  & \begin{pmatrix}
\mathbf{q}&
1
\end{pmatrix}
     \left(\begin{array}{c}
    -\Delta u \\
    -\Delta v \\
       (\Delta u \quad \Delta v)     \mathbf{q}^\top  
    \end{array}\right)=0.
\label{eq:plane}
\end{eqnarray}
According to Equation~\ref{eq:plane}, the vector $\begin{pmatrix} \mathbf{q} \quad 1 \end{pmatrix}$ is the normal of the plane formed by all the intersection points. We choose the normal pointing toward the image plane, which is
\begin{equation}
    \mathbf{n}' = -\left(
\mathbf{q} \quad 1 \right)^\top.
\end{equation}
Here, $\mathbf{n}'$ is a $3\times 1$ vector since $\mathbf{q}$ is a $1\times2$ vector. The denotation $'$ represents parameters in the ray space.

\noindent \textbf{Plane Normal in Camera Space.} With the normal direction derived in the ray space, we transform it back to the camera space by the local affine transformation as,
\begin{equation}
    \mathbf{n} = \mathbf{J}^\top \mathbf{n}',
\end{equation}
where $\mathbf{J}$ is the local affine matrix. After transformation, the vector $\mathbf{n}$ is normalized to unit length.

Our rasterized depth and normal maps are derived from general 3D Gaussian Splats under the local affine transformation assumption, which is assumed in the standard Gaussian Splatting \cite{kerbl20233d}. Consequently, our method can be directly integrated into existing methods utilizing 3D Gaussian Splatting.

\subsection{Loss Functions}
Even with our derived depth and normal maps, the Gaussian Splatting cannot recover shape details if it is only trained with the photometric supervision $\mathcal{L}_{c}$ that minimizes the difference between rendered and input images. 
To address this problem, we follow the 2D Gaussian Splatting~\cite{huang20242d} to apply additional depth distortion loss and normal consistency loss. We detail these two terms here to make this paper self-contained.

\noindent \textbf{Depth distortion loss.} The depth distortion loss encourages different Gaussian splats on a ray to be close to each other by minimizing the disparity of their depths as,
\begin{equation}
    \mathcal{L}_{d} = \sum_{i,j}\omega_i\omega_j(d_i-d_j)^2,
\end{equation}
where $\omega_i = \alpha_i \prod_{j=1}^{i-1}(1-\alpha_j)$ is the blending weight of $i-$th Gaussian and $d_i$ is its depth.

\noindent \textbf{Normal consistency loss.} The normal consistency loss ensures the Gaussian splats align with the surface by measuring the consistency between normal directions computed from the Gaussian and the depth map, respectively,
\begin{equation}
\mathcal{L}_{n} = \sum_{i} \omega_i (1-\bn_i^\top\tilde{\mathbf{n}}),
\end{equation}
where $\tilde{\mathbf{n}}$ is the surface normal direction obtained by applying finite-difference on the depth map.

Our final training loss $\mathcal{L}$ is,
\begin{equation}
    \mathcal{L} = \mathcal{L}_{c} + w_d \mathcal{L}_{d} + w_n \mathcal{L}_{n}.
\end{equation}

\section{Experiment}
We evaluate the performance of our method on both novel view synthesis and 3D reconstruction with standard benchmark datasets and compare it with state-of-the-art (SOTA) implicit and explicit approaches.
\setlength\tabcolsep{0.5em}
\begin{table*}[ht]
\centering
\caption{\textbf{Quantitative comparison on the DTU Dataset~\cite{jensen2014large}}. We report the Chamfer Distance error of different methods. Our method achieves the best performance among all explicit Gaussian Spaltting based methods, producing comparable accuracy as NeuraLangelo. We compare our method with other Gaussian Splatting methods under different resolutions. Methods with and without '(full resolution)' are trained under full resolution and half resolution respectively.}
\vspace{-0.2cm}
\resizebox{.98\textwidth}{!}{
\begin{tabular}{@{}llcccccccccccccccclcc}
\hline
 \multicolumn{3}{c}{} & 24 & 37 & 40 & 55 & 63 & 65 & 69 & 83 & 97 & 105 & 106 & 110 & 114 & 118 & 122 & & Mean & Time \\ \cline{4-18} \cline{20-21}
\multirow{4}{*}{\rotatebox[origin=c]{90}{implicit}} & NeRF~\cite{mildenhall2021nerf} & & 1.90 & 1.60 & 1.85 & 0.58 & 2.28 & 1.27 & 1.47 & 1.67 & 2.05 & 1.07 & 0.88 & 2.53 & 1.06 & 1.15 & 0.96 & & 1.49 & $>\text{12h}$\\
 & VolSDF~\cite{yariv2021volume} & &  1.14 &  1.26 &  0.81 & 0.49 & 1.25 &  \tbest0.70 &  \tbest0.72 &  1.29 & \tbest 1.18 &  \tbest 0.70 & 0.66 & 1.08 &  0.42 &  \tbest 0.61 &  0.55 & & 0.86 & $>\text{12h}$\\
 & NeuS~\cite{wang2021neus} & &  1.00 & 1.37 & 0.93 &  0.43 & 1.10 &  \sbest 0.65 &   \sbest 0.57 &  1.48 &  \sbest 1.09 &  0.83 &  \sbest 0.52 &  1.20 & \tbest 0.35 &  \sbest 0.49 &  0.54 & &  0.84 & $>\text{12h}$\\
 & Neuralangelo~\cite{li2023neuralangelo} & & \best 0.37 & \sbest 0.72 & \sbest 0.35 & \best 0.35 & \sbest 0.87 & \best 0.54 & \best 0.53 &  1.29 & \best 0.97 &  0.73 & \best 0.47 & \best 0.74 & \best 0.32 & \best 0.41 & \best 0.43 & & \best 0.61 & $>\text{12h}$\\ 
 \cline{2-2} \cline{4-18} \cline{20-21}
\multirow{9}{*}{\rotatebox[origin=c]{90}{explicit}} 
&  3D GS~\cite{kerbl20233d} & & 2.14 & 1.53 & 2.08 & 1.68 & 3.49 & 2.21 & 1.43 & 2.07 & 2.22 & 1.75 &  1.79 & 2.55 & 1.53 & 1.52 & 1.50 & & 1.96 & \sbest 5.2m\\
 &  SuGaR~\cite{guedon2023sugar} & & 1.47 & 1.33 & 1.13 & 0.61 & 2.25 & 1.71 & 1.15 & 1.63 & 1.62 & 1.07 & 0.79 & 2.45 & 0.98 & 0.88 & 0.79 & & 1.33  & 52m \\
 & 2D GS~\cite{huang20242d} &&  0.48 & 0.91 &  0.39 &  0.39 &  1.01 &  0.83 &  0.81 &  1.36 &  1.27 &  0.76  &  0.70 &  1.40 &   0.40 &   0.76 &  0.52 &&   0.80 & 8.9m \\
 & GOF~\cite{yu2024gaussian} & &  0.50 & 0.82 & 0.37 & \sbest 0.37 & 1.12 &  0.74 & 0.73 & \sbest 1.18 & 1.29 & \tbest 0.68 & 0.77 &  0.90 & 0.42 & 0.66 & 0.49 &&  0.74 & 55m\\
 & Our (20K iterations) & &  \tbest 0.45 & 0.74 & 0.36 &  0.39 & \tbest 0.90 &  0.73 & 0.77 & \best 1.13 & 1.26 &  \sbest 0.66 & \tbest 0.62 & \tbest 0.83 & 0.37 & \tbest 0.61 &  \tbest 0.46 && \tbest 0.69  & \best 5.0m\\
  & Our (30K iterations) & &  \tbest 0.46 &  \tbest 0.73 &  \tbest 0.33 &  \tbest 0.38 & \best 0.79 &  0.75 & 0.76 & \tbest 1.19 & 1.22 &  \best 0.62 & 0.70 & \sbest 0.78 & 0.36 & 0.68 &  0.47 && \sbest 0.68  & \tbest 8.3m\\
   \cline{2-2} \cline{4-18} \cline{20-21}
   & 2D GS (full resolution)~\cite{huang20242d} && 0.53  & 0.84 & 0.39  & \sbest 0.37  & 1.33  & 1.23  &  0.96 & 1.35  & 1.29  & 0.81  &  0.72 &  1.62 & 0.40  & 0.74  & 0.51  && 0.87   & 26.6m \\
 & GOF (full resolution)~\cite{yu2024gaussian} & &  0.52 & 0.85  & 0.50 & 0.40 & 1.38 & 1.05  & 0.96 & 1.31 & 1.35 & 0.80 & 0.82 & 0.86  & 0.47 & 0.62 & \tbest 0.46 && 0.82 & 136m\\
  & Our (full resolution) & & \sbest 0.40 & \best 0.71 & \best 0.33 & \sbest 0.37 & \sbest 0.87 &  0.79 & 0.77 &  1.22 & 1.26 & 0.70 & 0.65 & 0.85 & \sbest 0.33 & 0.66 & \sbest 0.44 && \tbest 0.69  & 20.2m\\
 \hline
\end{tabular}
}
\label{tab:dtu_result}
\vspace{-0.1cm}
\end{table*}

\begin{table}[t]
\centering
\caption{\textbf{Quantitative comparison on the Tanks \& Temples Dataset~\cite{Knapitsch2017}}. We report the F1-score and average optimization time. All results are evaluated with the official evaluation scripts.  $\sharp$ represents using marching tetrahedra algorithm\cite{yu2024gaussian} to extract meshes. Our method achieves the best F1 score among Gaussian Splatting and TSDF fusion based methods while maintaining the capacity of fast training.}
\vspace{-0.2cm}
\resizebox{0.98\columnwidth}{!}{
\begin{tabular}{@{}l|ccc|cccccccc}
 & \multicolumn{3}{c@{}|}{Implicit} & \multicolumn{3}{c@{}}{Explicit} \\ 
 & NeuS & Geo-Neus & Neurlangelo & SuGaR & 3D GS & 2D GS & GOF & $\text{GOF}^\sharp$ & Our &$\text{Our}^\sharp$\\ 
 \hline
Barn & 0.29 &  0.33 &  \best 0.70  & 0.14 & 0.13 & 0.36 & 0.37 & \sbest 0.51 &  \tbest 0.43 & \tbest 0.43\\
Caterpillar &  0.29 & 0.26 &  \sbest 0.36 & 0.16 & 0.08 & 0.23 & 0.21 & \best 0.41 & 0.26 & \tbest 0.32\\
Courthouse & \tbest 0.17 & 0.12 &  \best 0.28 & 0.08 & 0.09 & 0.13 & 0.11 & \best 0.28 & 0.11 & \sbest 0.21\\
Ignatius & \sbest 0.83 & 0.72 &  \best 0.89 & 0.33 & 0.04 & 0.44 & 0.63 & 0.68 & \tbest 0.73 & 0.69\\
Meetingroom &  0.24 & 0.20 &  \best 0.32 &  0.15 & 0.01 & 0.16 & 0.23 &\sbest 0.28 & 0.17 & \tbest 0.25\\
Truck &  0.45 &  0.45 &  0.48 &  0.26 & 0.19 & 0.26 &  0.50 & \best 0.59 & \sbest 0.53 & \tbest 0.51\\ 
\hline
Mean &  0.38 & 0.35 &  \best 0.50 & 0.19 & 0.09 & 0.30 & 0.34 & \sbest 0.46 & 0.37 & \tbest 0.40\\
Time & >24h & >24h & >24h & 73~m & \best 7.9~m & \tbest 12.3 ~m & 69~m & 69m & \sbest 11.5~m & \sbest 11.5~m\\ 
\end{tabular}
}
\label{tab:tnt}
\vspace{-0.2cm}
\end{table}

\subsection{Experimental Setup}

\subsubsection{Implementation Details}
We build our method upon the public code of 3D GS~\cite{kerbl20233d}, incorporating customized CUDA kernels for our rasterized depth, normal map computations, and regularization terms. 
We use the default parameters of 3D GS and incorporate the 3D filter proposed in Mip-Splatting~\cite{yu2023mip} and the densification proposed by GOF~\cite{yu2024gaussian}. 
Following the approach by GOF, we employ the decoupled appearance modeling from VastGaussian~\cite{Lin_2024_CVPR} on both the DTU~\cite{jensen2014large} and TNT~\cite{Knapitsch2017} datasets. We first train 3D Gaussians with only rendering losses for 15k iterations and then optimize the result further with our proposed depth and normal constraint for another 15k iterations.  
In all experiments, we set the weight parameters, $w_d=100$ and $w_n=5$, following the practice in GOF~\cite{yu2024gaussian}. Furthermore, we detach the gradient propagation of the blending weight $\omega$ when calculating depth distortion loss. All our experiments are conducted on a single NVIDIA H800 GPU.

\subsubsection{Mesh Extraction}
We render depth maps for all training views and construct a TSDF~\cite{curless1996volumetric} for mesh extraction with the Marching Cube~\cite{lorensen1998marching} algorithm.

\subsubsection{Datasets}
We conduct the surface reconstruction experiments on the DTU~\cite{jensen2014large} and Tanks \& Temples (TNT)~\cite{Knapitsch2017} datasets. 
Following the prior works, we select 15 scenes from the DTU dataset and 6 scenes from the TNT dataset for evaluation.
With the camera poses provided by these datasets, we apply COLMAP~\cite{schoenberger2016sfm} to generate a sparse point cloud of each scene for initialization. 

For novel view synthesis, we experiment on the Mip-NeRF360~\cite{barron2022mip} and the Synthetic-NeRF~\cite{mildenhall2021nerf} datasets. The Mip-NeRF360 dataset contains large indoor and outdoor scenes, while the Synthetic-NeRF contains object-level scenes with challenging reflections and detailed shapes.

\subsubsection{Evaluation protocols} To facilitate comparisons with previous methods, we report the \textbf{Chamfer Distance (CD)} on the DTU dataset and the \textbf{F1-score} on the TNT dataset. 
To evaluate the quality of novel view synthesis, we adopt \textbf{PSNR}, \textbf{SSIM}, and \textbf{LPIPS} as metrics.

\subsubsection{Baselines} We compare our method with SOTA Gaussian Splatting methods for surface reconstruction, including GOF~\cite{huang20242d}, SuGaR~\cite{guedon2023sugar}, 2D GS~\cite{huang20242d}, and 3D GS~\cite{kerbl20233d}. 
We also compare with NeRF-based implicit methods, including VolSDF~\cite{yariv2021volume}, NeuS~\cite{wang2021neus}, and Neuralangelo~\cite{li2023neuralangelo}. These methods adopt a Signed Distance Function (SDF) to represent the scene and transform the SDF to opacity for ray-tracing based volume rendering.

\begin{figure}[t]
    \centering
    \includegraphics[width=\linewidth]{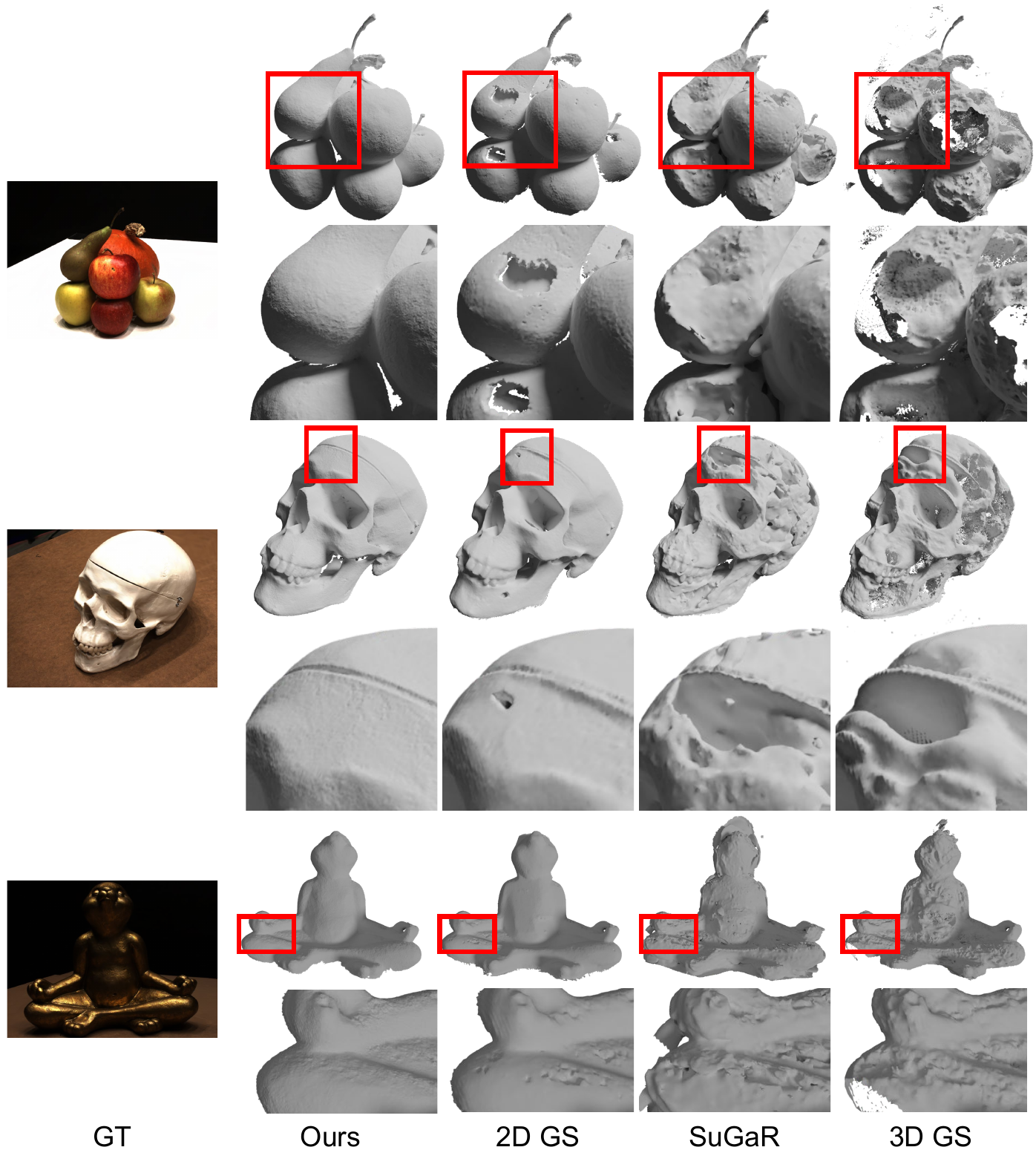}
    \caption{Qualitative comparison of our method and the previous Gaussian-based methods on the DTU dataset~\cite{jensen2014large}. 
    }
    \label{fig:our_dtu_reconstruction}
\end{figure}

\begin{figure*}[t]
    \centering
    \includegraphics[width=1\linewidth]{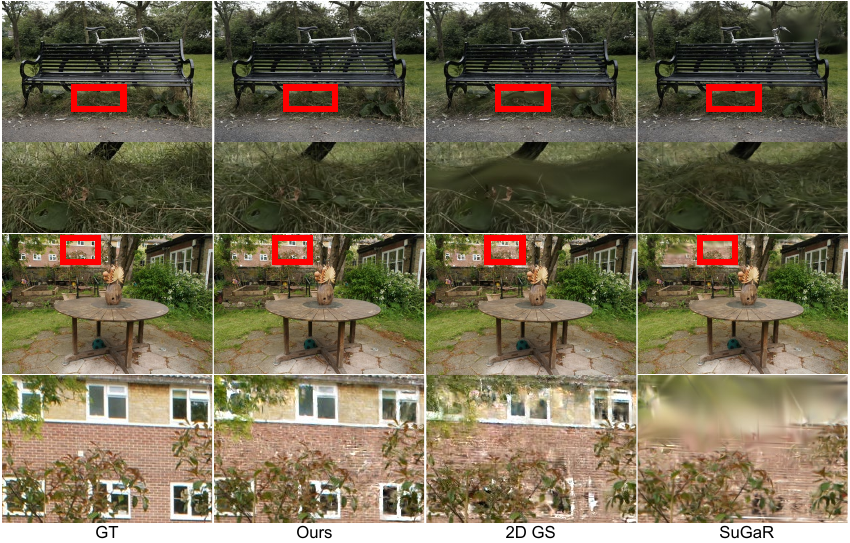}
    \caption{ Comparisons of our method to previous GS-based methods on novel view synthesis. Our method achieves high-quality images, 
    while others generate blurry results around the bench and artifacts at the distant wall, as shown in the framed regions.
    }
    \label{fig:ours_m360_nvs}
\end{figure*}

\begin{table}[t]
\centering
\caption{\textbf{Quantitative results on Mip-NeRF 360~\cite{barron2022mip} dataset.}
Our method achieved SOTA NVS results, especially in the outdoor scenes in terms of LPIPS.
}
\resizebox{0.98\columnwidth}{!}{
\begin{tabular}{@{}l|ccc|ccc}
 & \multicolumn{3}{c@{}|}{Outdoor Scene} & \multicolumn{3}{c@{}}{Indoor scene} \\ 
& PSNR~$\uparrow$ & SSIM~$\uparrow$ & LPIPS~$\downarrow$ & PSNR~$\uparrow$ & 
SSIM~$\uparrow$ & LIPPS~$\downarrow$ \\
\hline
NeRF & 21.46 & 0.458 & 0.515 & 26.84 &  0.790 & 0.370 \\
Deep Blending & 21.54 &0.524 & 0.364 & 26.40 & 0.844 & 0.261 \\
Instant NGP & 22.90 & 0.566 & 0.371 & 29.15 & 0.880 & 0.216 \\
MERF & 23.19 & 0.616 &   0.343 & 27.80 & 0.855 & 0.271 \\
MipNeRF360 &   24.47 &  0.691 &  0.283 &   \best 31.72 &  0.917 &  \sbest 0.180 \\
\hline
\hline
Mobile-NeRF & 21.95 & 0.470 & 0.470 & - & - & - \\
BakedSDF & 22.47 & 0.585 &  0.349 & 27.06 & 0.836 & 0.258 \\
SuGaR &  22.93 & 0.629 & 0.356 & 29.43 & 0.906 & 0.225 \\
BOG & 23.94 & 0.680 & 0.263 & 27.71 & 0873 & 0.227 \\
\hline
\hline
3D GS &   24.64 &   0.731 &  \tbest 0.234 &    30.41 &  0.920 &   0.189 \\
Mip-Splatting & \tbest 24.65 & \tbest 0.729 &  0.245 & \sbest 30.90 & \tbest 0.921 &  0.194  \\
2D GS &  24.34 & 0.717 & 0.246 & 30.40 & 0.916 & 0.195 \\
GOF & \sbest 24.82 & \sbest 0.750 &\sbest  0.202 & \tbest 30.79 & \sbest 0.924 &\tbest  0.184 \\
Ours & \best 25.17 & \best 0.764 &\best  0.199 & 30.74 &  \best 0.928 &  \best 0.165
\end{tabular}
}
\label{tab:mipnerf360}
\end{table}
\begin{table*}[h]
\centering
\caption{\textbf{PSNR scores for Synthetic NeRF \cite{mildenhall2021nerf}}. Our method outperforms others in novel view synthesis. We use $\star$ to represent 2D GS trained with regularization. To ensure fair comparisons, we train our model, GOF, and 2D GS with regularization. While 2D GS reports geometry results with regularization, it does not do so for novel view synthesis.}
\label{tab:nerf_synthetic}
\resizebox{0.8\linewidth}{!}{
\begin{tabular}{l|ccccccccc}
 & \textbf{Mic} & \textbf{Chair} & \textbf{Ship} & \textbf{Materials} & \textbf{Lego} & \textbf{Drums} & \textbf{Ficus} & \textbf{Hotdog} & \textbf{Avg.} \\
\hline
Plenoxels   & 33.26         & 33.98         & 29.62         & 29.14         & 34.10         & 25.35         & 31.83         & 36.81         & 31.76 \\
INGP-Base   & \sbest 36.22  & 35.00         & \sbest 31.10  & 29.78         & \best 36.39   & 26.02         & 33.51         & 37.40         & 33.18 \\
Mip-NeRF    & \best 36.51   & 35.14         & 30.41         & \best 30.71   & 35.70         & 25.48         & 33.29         & 37.48         & 33.09 \\
Point-NeRF  & \tbest 35.95  & 35.40         & \tbest 30.97  & 29.61         & 35.04         & 26.06         & \best 36.13   & 37.30         & 33.30 \\
3D GS        & 35.36         & \tbest35.83   & 30.80         & 30.00         & \sbest 35.78  & \tbest 26.15  & 34.87         & \sbest 37.72   & \tbest 33.32 \\
$\text{2D GS}^{\star}$       &  34.86        &   34.77       &   30.47        &   29.58       &   33.98       &   25.47      &    35.25     &      36.83    &  32.65\\
2D GS &35.09       & 35.05         & 30.60         & 29.74         & 35.10         & 26.05         & \sbest35.57   & 37.36         & 33.07 \\
GOF         & 35.81         & \sbest36.34   & 30.71         & \sbest30.19   & 35.64         & \sbest 26.21  & 35.25         & \tbest37.54   & \sbest33.46\\
Our         & 35.46         & \best 36.46   & \best 31.35   & \tbest 30.34  & \tbest 35.75  & \best 26.29   & \tbest 35.37  & \best 37.76   & \best 33.60\\
\end{tabular}
}
\label{tab:nerf}
\end{table*}

\subsection{Comparison}
\subsubsection{Surface Reconstruction}

We compare our method with existing methods on the DTU and TNT datasets in Table~\ref{tab:dtu_result} and Table~\ref{tab:tnt}, respectively.
As shown in Table~\ref{tab:dtu_result}, our method produces a mean Chamfer Distance (CD) error of $0.68$mm, outperforming all Gaussian splatting-based methods. This result is competitive with Neuralangelo, which achieves $0.61$mm mean CD error. Please note that all Gaussian Splatting based methods are evaluated using half-image resolution. We further include the comparison with 2D GS and GOF on full-resolution images at the bottom of Table~\ref{tab:dtu_result}.

\begin{figure*}[t]
    \centering
    \setlength{\abovecaptionskip}{0.cm}
    \includegraphics[width=0.9\linewidth]{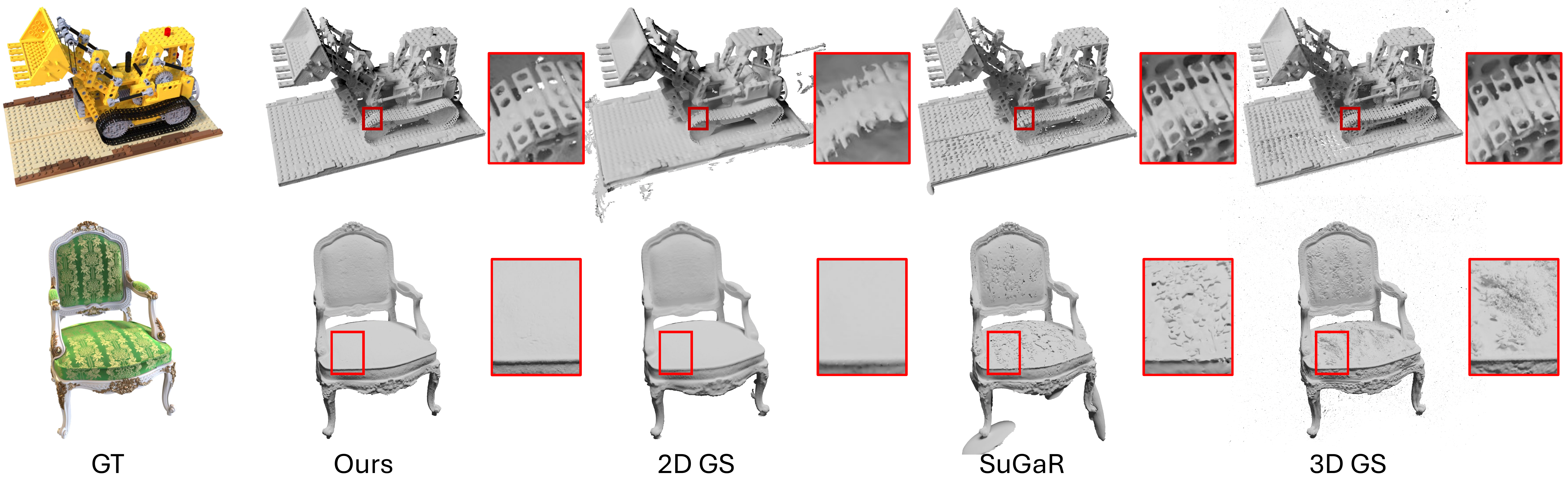}
    \caption{Qualitative comparisons of our method and previous Gaussian-based methods on the Synthetic NeRF dataset~\cite{mildenhall2021nerf}. 
    }
    \label{fig:ours_nerf_reconstruction}
\end{figure*}

\begin{figure*}[h]
    \centering
    \setlength{\abovecaptionskip}{0.cm}
    \includegraphics[width=\linewidth]{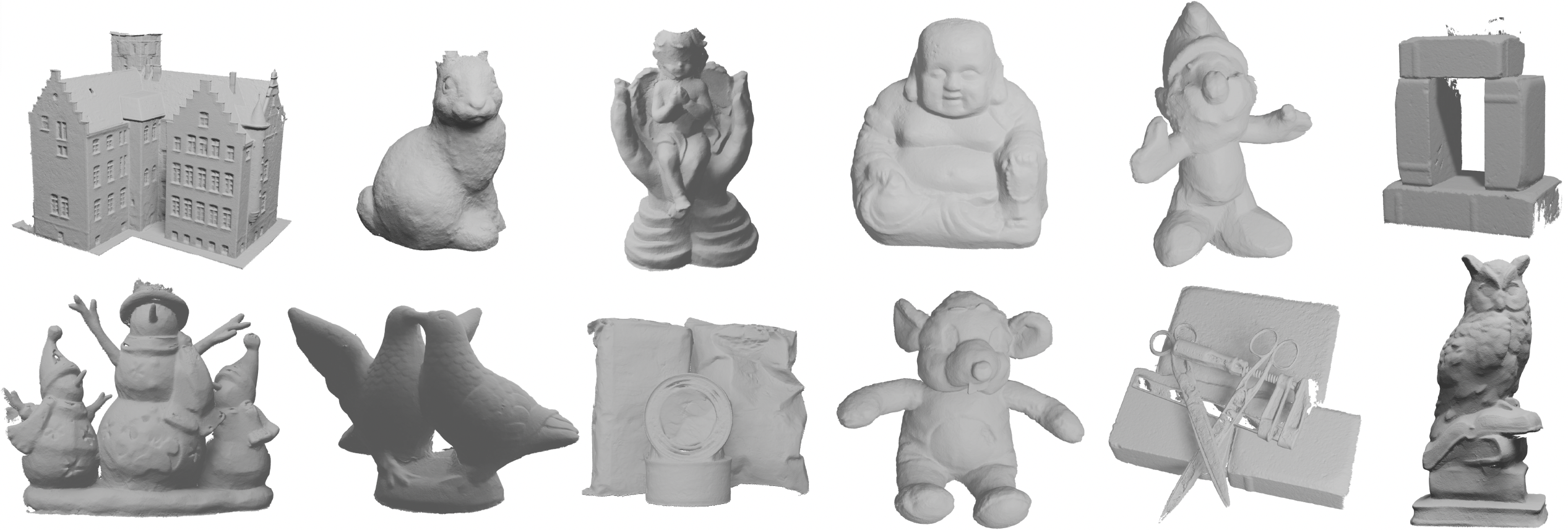}
    \caption{Surface reconstruction results on the DTU dataset~\cite{jensen2014large}.}
    \label{fig:our_dtu}
    \vspace{-0.1in}
\end{figure*}

Figure~\ref{fig:our_dtu_reconstruction} offers qualitative comparisons among different GS-based methods on the DTU dataset. Our method consistently produces smooth and precise shapes, while 3D GS tends to generate noisy meshes due to imprecise depth approximation. At the same time, 2D GS and SuGaR exhibit instability in areas with specular highlights, leading to inaccurate surface reconstructions. In comparison, our method is more robust to these problems. Additional qualitative comparisons can be found in Figure~\ref{fig:ours_nerf_reconstruction}. More examples of surface meshes reconstructed by our method are visualized in Figure~\ref{fig:our_dtu}. 

\begin{figure*}[t]
    \centering
    \setlength{\abovecaptionskip}{0.cm}
    \includegraphics[width=\linewidth]{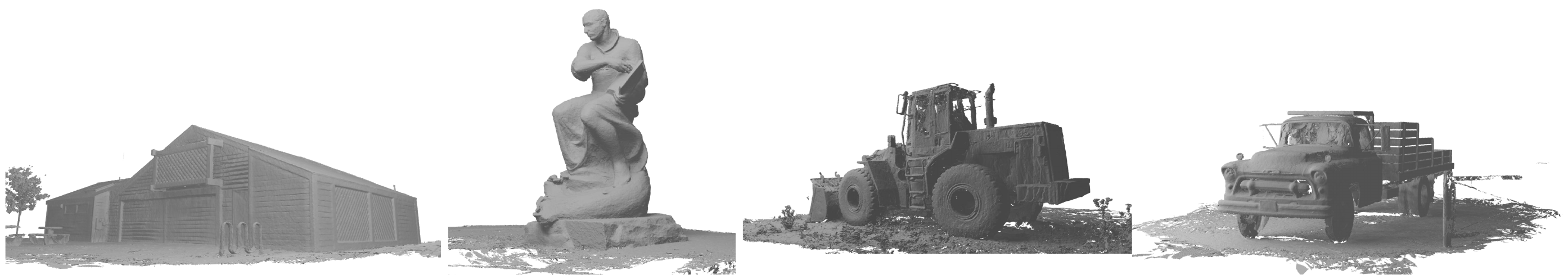}
    \caption{Surface reconstruction results on the Tanks \& Templs dataset~\cite{Knapitsch2017}.}
    \label{fig:our_tnt}
\end{figure*}

Table~\ref{tab:tnt} compares our method with other methods on the TNT dataset. Again, our method outperforms most of the GS-based methods. However, our F1-score is lower than Neuralangelo and close to NeuS. This is mainly due to the voxel resolution during our TSDF fusion of the depth maps. The TNT dataset contains large-scale scenes such as the `\emph{Courthouse}' and `\emph{Meetingroom}'. Due to memory constraints, our current version is limited to a low-resolution voxel grid in TSDF fusion. We could employ a multi-resolution TSDF fusion~\cite{vespa2019adaptive} to obtain a better F1 score. We leave this task to future research. GOF produces the highest F1 score among all GS-based methods, benefiting from its Marching Tetrahedra for surface extraction, which is not limited by low-resolution voxels when dealing with large-scale scenes. 
Figure~\ref{fig:our_tnt} visualizes some of the reconstructed meshes by our method.

\begin{figure*}[t]
    \centering
    \setlength{\abovecaptionskip}{0.cm}
    \includegraphics[width=0.91\linewidth]{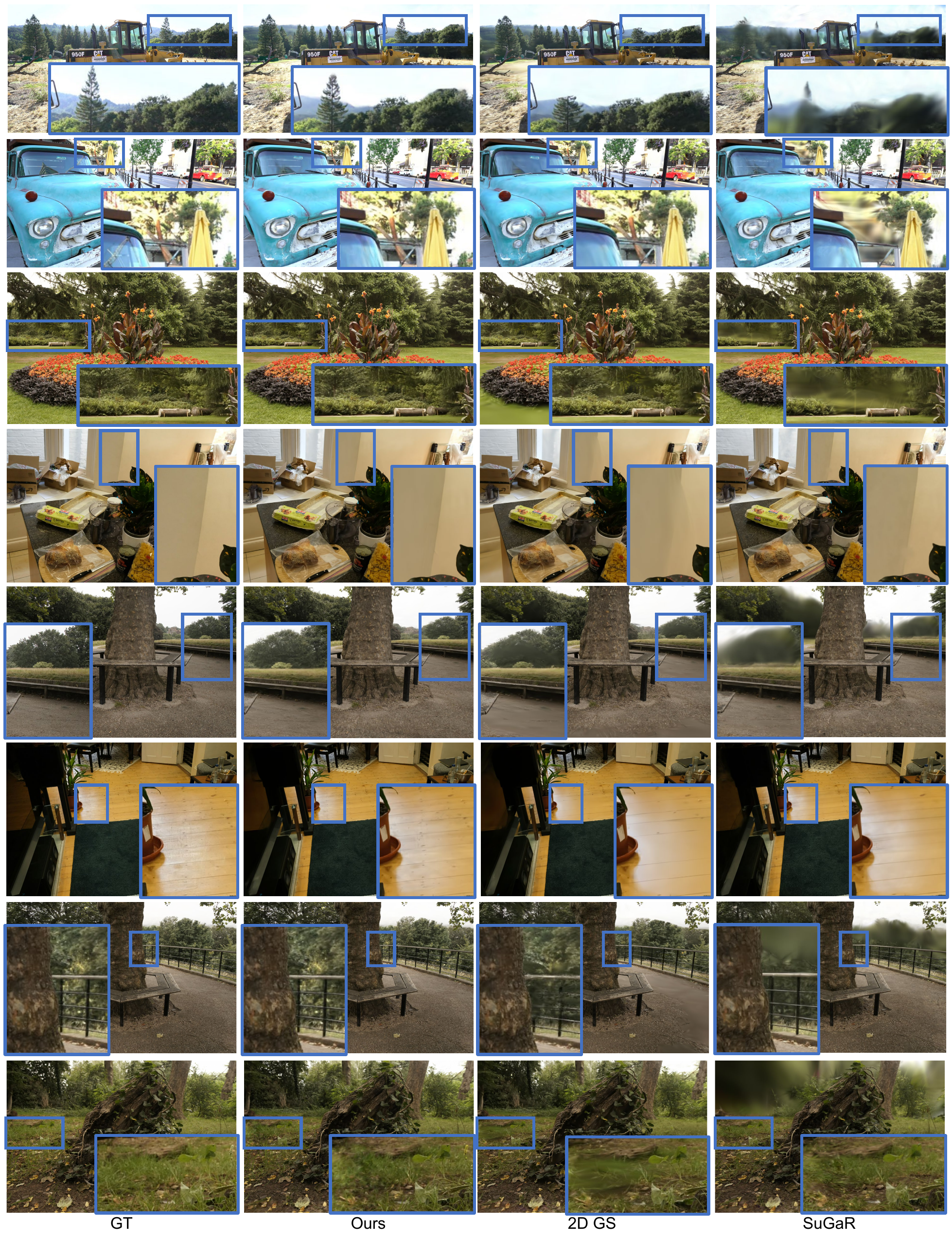}
    \caption{Additional comparisons on novel view synthesis between our and previous methods. The testing scenes are from the Tanks \& Temples~\cite{Knapitsch2017} and Mip-NeRF360~\cite{barron2022mip} datasets.
    Please zoom in on the digital version to see details in the framed regions.}
    \label{fig:our_nvs}
    \vspace{-0.5in}
\end{figure*}

\subsubsection{Novel view synthesis}
We further compare our method against previous methods on the Mip-NeRF360 and Synthetic NeRF datasets to evaluate their novel view synthesis capability. The quantitative results presented in Table~\ref{tab:mipnerf360} and Table~\ref{tab:nerf} show that our method achieves the highest average PSNR on the Synthetic NeRF dataset and excels in most metrics on the Mip-NeRF360 dataset. 
In comparison, the 2D GS and SuGaR methods generate inferior novel view rendering than the standard 3D GS. This outcome indicates that the planar Gaussian constraints adopted in 2D GS and SuGaR hurt the model's performance in complex scenes. Our method, which retains the original 3D GS, achieves better data representation and more effective novel view synthesis results.

We show the novel view synthesis results on the Mip-NeRF360 dataset in Figure~\ref{fig:ours_m360_nvs}.
As shown in the framed regions, our method can render high-quality images, while SuGaR generates blurry results for places with complicated shapes, such as grass and leaves. Due to unstable optimization, it also generates artifacts at the distant wall. Please zoom in on the digital version for clearer visualization.
More qualitative comparisons on novel view synthesis can be found in Figure~\ref{fig:our_nvs}.

\subsubsection{Computational Efficiency}
Table~\ref{tab:dtu_result} and Table~\ref{tab:tnt} provide the average training time of different methods. GS-based explicit methods are significantly more computationally efficient than implicit NeRF-based methods. Our method reconstructs a scene in about $8.3$ minutes on the DTU dataset and $11.5$ minutes on the TNT dataset, while all implicit methods take hours. As shown in Table~\ref{tab:dtu_result}, 2D GS takes slightly longer, at 8.9 minutes on the DTU dataset and 12.3 minutes on the TNT dataset. GOF takes about one hour for training, which can be attributed to its time-consuming ray tracing approach for depth estimation. Compared with 3D GS, our additional computation time mainly stems from the forward and backward process of pixel-wise geometric regularization terms.
We further test our method with different numbers of iterations on the DTU dataset (shown in Table~\ref{tab:dtu_result}). Our method achieves better accuracy and efficiency than 3D GS with fewer optimization steps, i.e., at 20k iterations.

\section{Conclusion}
We introduce RaDe-GS, which includes a rasterized method to compute the depth and surface normal maps, to improve Gaussian Splatting's shape reconstruction capability while maintaining its training and rendering efficiency. Our method is derived from general Gaussian primitives, it can be easily integrated into any Gaussian-Splatting-based method.
Our experiments on multiple public datasets demonstrate that our RaDe-GS produces superior results than previous implicit or explicit methods.

\noindent \textbf{Limitation.} Our current TSDF fusion is limited to low-resolution voxel grids for large-scale scenes, thus hindering accurate surface extraction from Gaussians. 
The adaptive TSDF fusion techniques can be used to extract mesh in a coarse-to-fine manner, which leads to less memory consumption but better geometry quality.
For the object containing reflective surfaces, e.g, the Metal Scissor of DTU dataset (shown in Figure~\ref{fig:our_dtu}), our method fails to accurately reconstruct the reflective surfaces limited by the simple color function used in 3D GS.
This problem may be alleviated by combining with advanced color representation proposed in GaussianShader~\cite{jiang2023gaussianshader}.

\setcitestyle{square}
\bibliographystyle{ACM-Reference-Format}
\bibliography{sample-bibliography}


\begin{thebibliography}{51}


\ifx \showCODEN    \undefined \def \showCODEN     #1{\unskip}     \fi
\ifx \showDOI      \undefined \def \showDOI       #1{#1}\fi
\ifx \showISBNx    \undefined \def \showISBNx     #1{\unskip}     \fi
\ifx \showISBNxiii \undefined \def \showISBNxiii  #1{\unskip}     \fi
\ifx \showISSN     \undefined \def \showISSN      #1{\unskip}     \fi
\ifx \showLCCN     \undefined \def \showLCCN      #1{\unskip}     \fi
\ifx \shownote     \undefined \def \shownote      #1{#1}          \fi
\ifx \showarticletitle \undefined \def \showarticletitle #1{#1}   \fi
\ifx \showURL      \undefined \def \showURL       {\relax}        \fi
\providecommand\bibfield[2]{#2}
\providecommand\bibinfo[2]{#2}
\providecommand\natexlab[1]{#1}
\providecommand\showeprint[2][]{arXiv:#2}

\bibitem[Barnes et~al\mbox{.}(2009)]%
        {barnes2009patchmatch}
\bibfield{author}{\bibinfo{person}{Connelly Barnes}, \bibinfo{person}{Eli Shechtman}, \bibinfo{person}{Adam Finkelstein}, {and} \bibinfo{person}{Dan~B Goldman}.} \bibinfo{year}{2009}\natexlab{}.
\newblock \showarticletitle{PatchMatch: A randomized correspondence algorithm for structural image editing}.
\newblock \bibinfo{journal}{\emph{ACM Trans. Graph.}} \bibinfo{volume}{28}, \bibinfo{number}{3} (\bibinfo{year}{2009}), \bibinfo{pages}{24}.
\newblock


\bibitem[Barron et~al\mbox{.}(2021)]%
        {barron2021mip}
\bibfield{author}{\bibinfo{person}{Jonathan~T Barron}, \bibinfo{person}{Ben Mildenhall}, \bibinfo{person}{Matthew Tancik}, \bibinfo{person}{Peter Hedman}, \bibinfo{person}{Ricardo Martin-Brualla}, {and} \bibinfo{person}{Pratul~P Srinivasan}.} \bibinfo{year}{2021}\natexlab{}.
\newblock \showarticletitle{Mip-nerf: A multiscale representation for anti-aliasing neural radiance fields}. In \bibinfo{booktitle}{\emph{Proceedings of the IEEE/CVF International Conference on Computer Vision}}. \bibinfo{pages}{5855--5864}.
\newblock


\bibitem[Barron et~al\mbox{.}(2022)]%
        {barron2022mip}
\bibfield{author}{\bibinfo{person}{Jonathan~T Barron}, \bibinfo{person}{Ben Mildenhall}, \bibinfo{person}{Dor Verbin}, \bibinfo{person}{Pratul~P Srinivasan}, {and} \bibinfo{person}{Peter Hedman}.} \bibinfo{year}{2022}\natexlab{}.
\newblock \showarticletitle{Mip-nerf 360: Unbounded anti-aliased neural radiance fields}. In \bibinfo{booktitle}{\emph{Proceedings of the IEEE/CVF Conference on Computer Vision and Pattern Recognition}}. \bibinfo{pages}{5470--5479}.
\newblock


\bibitem[Bleyer et~al\mbox{.}(2011)]%
        {Bleyer2011}
\bibfield{author}{\bibinfo{person}{Michael Bleyer}, \bibinfo{person}{Christoph Rhemann}, {and} \bibinfo{person}{Carsten Rother}.} \bibinfo{year}{2011}\natexlab{}.
\newblock \showarticletitle{PatchMatch Stereo - Stereo Matching with Slanted Support Windows}. In \bibinfo{booktitle}{\emph{BMVC}}.
\newblock


\bibitem[Chen et~al\mbox{.}(2023b)]%
        {chen2023neusg}
\bibfield{author}{\bibinfo{person}{Hanlin Chen}, \bibinfo{person}{Chen Li}, {and} \bibinfo{person}{Gim~Hee Lee}.} \bibinfo{year}{2023}\natexlab{b}.
\newblock \showarticletitle{Neusg: Neural implicit surface reconstruction with 3d gaussian splatting guidance}.
\newblock \bibinfo{journal}{\emph{arXiv preprint arXiv:2312.00846}} (\bibinfo{year}{2023}).
\newblock


\bibitem[Chen et~al\mbox{.}(2023a)]%
        {chen2023mobilenerf}
\bibfield{author}{\bibinfo{person}{Zhiqin Chen}, \bibinfo{person}{Thomas Funkhouser}, \bibinfo{person}{Peter Hedman}, {and} \bibinfo{person}{Andrea Tagliasacchi}.} \bibinfo{year}{2023}\natexlab{a}.
\newblock \showarticletitle{Mobilenerf: Exploiting the polygon rasterization pipeline for efficient neural field rendering on mobile architectures}. In \bibinfo{booktitle}{\emph{Proceedings of the IEEE/CVF Conference on Computer Vision and Pattern Recognition}}. \bibinfo{pages}{16569--16578}.
\newblock


\bibitem[Cheng et~al\mbox{.}(2024)]%
        {cheng2024gaussianpro}
\bibfield{author}{\bibinfo{person}{Kai Cheng}, \bibinfo{person}{Xiaoxiao Long}, \bibinfo{person}{Kaizhi Yang}, \bibinfo{person}{Yao Yao}, \bibinfo{person}{Wei Yin}, \bibinfo{person}{Yuexin Ma}, \bibinfo{person}{Wenping Wang}, {and} \bibinfo{person}{Xuejin Chen}.} \bibinfo{year}{2024}\natexlab{}.
\newblock \showarticletitle{GaussianPro: 3D Gaussian Splatting with Progressive Propagation}.
\newblock \bibinfo{journal}{\emph{arXiv preprint arXiv:2402.14650}} (\bibinfo{year}{2024}).
\newblock


\bibitem[Collins(1996)]%
        {Collins1996}
\bibfield{author}{\bibinfo{person}{Robert Collins}.} \bibinfo{year}{1996}\natexlab{}.
\newblock \showarticletitle{A space-sweep approach to true multi-image matching}. In \bibinfo{booktitle}{\emph{Conference on Computer Vision and Pattern Recognition (CVPR)}}.
\newblock


\bibitem[Curless and Levoy(1996a)]%
        {Curless1996}
\bibfield{author}{\bibinfo{person}{Brian Curless} {and} \bibinfo{person}{Marc Levoy}.} \bibinfo{year}{1996}\natexlab{a}.
\newblock \showarticletitle{A Volumetric Method for Building Complex Models from Range Images}. In \bibinfo{booktitle}{\emph{SIGGRAPH}}.
\newblock


\bibitem[Curless and Levoy(1996b)]%
        {curless1996volumetric}
\bibfield{author}{\bibinfo{person}{Brian Curless} {and} \bibinfo{person}{Marc Levoy}.} \bibinfo{year}{1996}\natexlab{b}.
\newblock \showarticletitle{A volumetric method for building complex models from range images}. In \bibinfo{booktitle}{\emph{Proceedings of the 23rd annual conference on Computer graphics and interactive techniques}}. \bibinfo{pages}{303--312}.
\newblock


\bibitem[Ding et~al\mbox{.}(2022)]%
        {Ding2022}
\bibfield{author}{\bibinfo{person}{Yikang Ding}, \bibinfo{person}{Wentao Yuan}, \bibinfo{person}{Qingtian Zhu}, \bibinfo{person}{Haotian Zhang}, \bibinfo{person}{Xiangyue Liu}, \bibinfo{person}{Yuanjiang Wang}, {and} \bibinfo{person}{Xiao Liu}.} \bibinfo{year}{2022}\natexlab{}.
\newblock \showarticletitle{TransMVSNet: Global Context-aware Multi-view Stereo Network with Transformers}. In \bibinfo{booktitle}{\emph{Conference on Computer Vision and Pattern Recognition (CVPR)}}.
\newblock


\bibitem[Fan et~al\mbox{.}(2023)]%
        {fan2023lightgaussian}
\bibfield{author}{\bibinfo{person}{Zhiwen Fan}, \bibinfo{person}{Kevin Wang}, \bibinfo{person}{Kairun Wen}, \bibinfo{person}{Zehao Zhu}, \bibinfo{person}{Dejia Xu}, {and} \bibinfo{person}{Zhangyang Wang}.} \bibinfo{year}{2023}\natexlab{}.
\newblock \showarticletitle{Lightgaussian: Unbounded 3d gaussian compression with 15x reduction and 200+ fps}.
\newblock \bibinfo{journal}{\emph{arXiv preprint arXiv:2311.17245}} (\bibinfo{year}{2023}).
\newblock


\bibitem[Fridovich-Keil et~al\mbox{.}(2022)]%
        {fridovich2022plenoxels}
\bibfield{author}{\bibinfo{person}{Sara Fridovich-Keil}, \bibinfo{person}{Alex Yu}, \bibinfo{person}{Matthew Tancik}, \bibinfo{person}{Qinhong Chen}, \bibinfo{person}{Benjamin Recht}, {and} \bibinfo{person}{Angjoo Kanazawa}.} \bibinfo{year}{2022}\natexlab{}.
\newblock \showarticletitle{Plenoxels: Radiance fields without neural networks}. In \bibinfo{booktitle}{\emph{Proceedings of the IEEE/CVF Conference on Computer Vision and Pattern Recognition}}. \bibinfo{pages}{5501--5510}.
\newblock


\bibitem[Fuhrmann et~al\mbox{.}(2014)]%
        {fuhrmann2014mve}
\bibfield{author}{\bibinfo{person}{Simon Fuhrmann}, \bibinfo{person}{Fabian Langguth}, {and} \bibinfo{person}{Michael Goesele}.} \bibinfo{year}{2014}\natexlab{}.
\newblock \showarticletitle{Mve-a multi-view reconstruction environment.}
\newblock \bibinfo{journal}{\emph{GCH}}  \bibinfo{volume}{3} (\bibinfo{year}{2014}), \bibinfo{pages}{4}.
\newblock


\bibitem[Gu et~al\mbox{.}(2019)]%
        {gu2019cas}
\bibfield{author}{\bibinfo{person}{Xiaodong Gu}, \bibinfo{person}{Zhiwen Fan}, \bibinfo{person}{Siyu Zhu}, \bibinfo{person}{Zuozhuo Dai}, \bibinfo{person}{Feitong Tan}, {and} \bibinfo{person}{Ping Tan}.} \bibinfo{year}{2019}\natexlab{}.
\newblock \showarticletitle{Cascade Cost Volume for High-Resolution Multi-View Stereo and Stereo Matching}.
\newblock \bibinfo{journal}{\emph{arxiv preprint arXiv:1912.06378}}.
\newblock


\bibitem[Gu et~al\mbox{.}(2020)]%
        {gu2020cascade}
\bibfield{author}{\bibinfo{person}{Xiaodong Gu}, \bibinfo{person}{Zhiwen Fan}, \bibinfo{person}{Siyu Zhu}, \bibinfo{person}{Zuozhuo Dai}, \bibinfo{person}{Feitong Tan}, {and} \bibinfo{person}{Ping Tan}.} \bibinfo{year}{2020}\natexlab{}.
\newblock \showarticletitle{Cascade cost volume for high-resolution multi-view stereo and stereo matching}. In \bibinfo{booktitle}{\emph{Conference on Computer Vision and Pattern Recognition (CVPR)}}. \bibinfo{pages}{2495--2504}.
\newblock


\bibitem[Gu{\'e}don and Lepetit(2023)]%
        {guedon2023sugar}
\bibfield{author}{\bibinfo{person}{Antoine Gu{\'e}don} {and} \bibinfo{person}{Vincent Lepetit}.} \bibinfo{year}{2023}\natexlab{}.
\newblock \showarticletitle{SuGaR: Surface-aligned gaussian splatting for efficient 3d mesh reconstruction and high-quality mesh rendering}.
\newblock \bibinfo{journal}{\emph{arXiv preprint arXiv:2311.12775}} (\bibinfo{year}{2023}).
\newblock


\bibitem[Hedman et~al\mbox{.}(2021)]%
        {hedman2021baking}
\bibfield{author}{\bibinfo{person}{Peter Hedman}, \bibinfo{person}{Pratul~P Srinivasan}, \bibinfo{person}{Ben Mildenhall}, \bibinfo{person}{Jonathan~T Barron}, {and} \bibinfo{person}{Paul Debevec}.} \bibinfo{year}{2021}\natexlab{}.
\newblock \showarticletitle{Baking neural radiance fields for real-time view synthesis}. In \bibinfo{booktitle}{\emph{Proceedings of the IEEE/CVF International Conference on Computer Vision}}. \bibinfo{pages}{5875--5884}.
\newblock


\bibitem[Huang et~al\mbox{.}(2024)]%
        {huang20242d}
\bibfield{author}{\bibinfo{person}{Binbin Huang}, \bibinfo{person}{Zehao Yu}, \bibinfo{person}{Anpei Chen}, \bibinfo{person}{Andreas Geiger}, {and} \bibinfo{person}{Shenghua Gao}.} \bibinfo{year}{2024}\natexlab{}.
\newblock \showarticletitle{2D Gaussian Splatting for Geometrically Accurate Radiance Fields}.
\newblock \bibinfo{journal}{\emph{arXiv preprint arXiv:2403.17888}} (\bibinfo{year}{2024}).
\newblock


\bibitem[Jensen et~al\mbox{.}(2014)]%
        {jensen2014large}
\bibfield{author}{\bibinfo{person}{Rasmus Jensen}, \bibinfo{person}{Anders Dahl}, \bibinfo{person}{George Vogiatzis}, \bibinfo{person}{Engin Tola}, {and} \bibinfo{person}{Henrik Aan{\ae}s}.} \bibinfo{year}{2014}\natexlab{}.
\newblock \showarticletitle{Large scale multi-view stereopsis evaluation}. In \bibinfo{booktitle}{\emph{Proceedings of the IEEE conference on computer vision and pattern recognition}}. \bibinfo{pages}{406--413}.
\newblock


\bibitem[Jiang et~al\mbox{.}(2023)]%
        {jiang2023gaussianshader}
\bibfield{author}{\bibinfo{person}{Yingwenqi Jiang}, \bibinfo{person}{Jiadong Tu}, \bibinfo{person}{Yuan Liu}, \bibinfo{person}{Xifeng Gao}, \bibinfo{person}{Xiaoxiao Long}, \bibinfo{person}{Wenping Wang}, {and} \bibinfo{person}{Yuexin Ma}.} \bibinfo{year}{2023}\natexlab{}.
\newblock \showarticletitle{GaussianShader: 3D Gaussian Splatting with Shading Functions for Reflective Surfaces}.
\newblock \bibinfo{journal}{\emph{arXiv preprint arXiv:2311.17977}} (\bibinfo{year}{2023}).
\newblock


\bibitem[Kendall et~al\mbox{.}(2017)]%
        {Kendall2017}
\bibfield{author}{\bibinfo{person}{Alex Kendall}, \bibinfo{person}{Hayk Martirosyan}, \bibinfo{person}{Saumitro Dasgupta}, \bibinfo{person}{Peter Henry}, \bibinfo{person}{Ryan Kennedy}, \bibinfo{person}{Abraham Bachrach}, {and} \bibinfo{person}{Adam Bry}.} \bibinfo{year}{2017}\natexlab{}.
\newblock \showarticletitle{End-to-End Learning of Geometry and Context for Deep Stereo Regression}. In \bibinfo{booktitle}{\emph{Conference on Computer Vision and Pattern Recognition (CVPR)}}.
\newblock


\bibitem[Kerbl et~al\mbox{.}(2023)]%
        {kerbl20233d}
\bibfield{author}{\bibinfo{person}{Bernhard Kerbl}, \bibinfo{person}{Georgios Kopanas}, \bibinfo{person}{Thomas Leimk{\"u}hler}, {and} \bibinfo{person}{George Drettakis}.} \bibinfo{year}{2023}\natexlab{}.
\newblock \showarticletitle{3d gaussian splatting for real-time radiance field rendering}.
\newblock \bibinfo{journal}{\emph{ACM Transactions on Graphics}} \bibinfo{volume}{42}, \bibinfo{number}{4} (\bibinfo{year}{2023}), \bibinfo{pages}{1--14}.
\newblock


\bibitem[Knapitsch et~al\mbox{.}(2017)]%
        {Knapitsch2017}
\bibfield{author}{\bibinfo{person}{Arno Knapitsch}, \bibinfo{person}{Jaesik Park}, \bibinfo{person}{Qian-Yi Zhou}, {and} \bibinfo{person}{Vladlen Koltun}.} \bibinfo{year}{2017}\natexlab{}.
\newblock \showarticletitle{Tanks and Temples: Benchmarking Large-Scale Scene Reconstruction}.
\newblock \bibinfo{journal}{\emph{ACM Transactions on Graphics}} \bibinfo{volume}{36}, \bibinfo{number}{4} (\bibinfo{year}{2017}).
\newblock


\bibitem[Kolmogorov and Zabih(2001)]%
        {Kolmogorov2001}
\bibfield{author}{\bibinfo{person}{Vladimir Kolmogorov} {and} \bibinfo{person}{Ramin Zabih}.} \bibinfo{year}{2001}\natexlab{}.
\newblock \showarticletitle{Computing visual correspondence with occlusions using graph cuts}. In \bibinfo{booktitle}{\emph{ICCV}}. IEEE.
\newblock


\bibitem[Kolmogorov and Zabin(2004)]%
        {kolmogorov2004energy}
\bibfield{author}{\bibinfo{person}{Vladimir Kolmogorov} {and} \bibinfo{person}{Ramin Zabin}.} \bibinfo{year}{2004}\natexlab{}.
\newblock \showarticletitle{What energy functions can be minimized via graph cuts?}
\newblock \bibinfo{journal}{\emph{IEEE transactions on pattern analysis and machine intelligence}} \bibinfo{volume}{26}, \bibinfo{number}{2} (\bibinfo{year}{2004}), \bibinfo{pages}{147--159}.
\newblock


\bibitem[Li et~al\mbox{.}(2023)]%
        {li2023neuralangelo}
\bibfield{author}{\bibinfo{person}{Zhaoshuo Li}, \bibinfo{person}{Thomas Müller}, \bibinfo{person}{Alex Evans}, \bibinfo{person}{Russell~H. Taylor}, \bibinfo{person}{Mathias Unberath}, \bibinfo{person}{Ming-Yu Liu}, {and} \bibinfo{person}{Chen-Hsuan Lin}.} \bibinfo{year}{2023}\natexlab{}.
\newblock \bibinfo{title}{Neuralangelo: High-Fidelity Neural Surface Reconstruction}.
\newblock
\newblock
\showeprint[arxiv]{2306.03092}~[cs.CV]


\bibitem[Lin et~al\mbox{.}(2024b)]%
        {lin2024vastgaussian}
\bibfield{author}{\bibinfo{person}{Jiaqi Lin}, \bibinfo{person}{Zhihao Li}, \bibinfo{person}{Xiao Tang}, \bibinfo{person}{Jianzhuang Liu}, \bibinfo{person}{Shiyong Liu}, \bibinfo{person}{Jiayue Liu}, \bibinfo{person}{Yangdi Lu}, \bibinfo{person}{Xiaofei Wu}, \bibinfo{person}{Songcen Xu}, \bibinfo{person}{Youliang Yan}, {et~al\mbox{.}}} \bibinfo{year}{2024}\natexlab{b}.
\newblock \showarticletitle{VastGaussian: Vast 3D Gaussians for Large Scene Reconstruction}.
\newblock \bibinfo{journal}{\emph{arXiv preprint arXiv:2402.17427}} (\bibinfo{year}{2024}).
\newblock


\bibitem[Lin et~al\mbox{.}(2024a)]%
        {Lin_2024_CVPR}
\bibfield{author}{\bibinfo{person}{Jiaqi Lin}, \bibinfo{person}{Zhihao Li}, \bibinfo{person}{Xiao Tang}, \bibinfo{person}{Jianzhuang Liu}, \bibinfo{person}{Shiyong Liu}, \bibinfo{person}{Jiayue Liu}, \bibinfo{person}{Yangdi Lu}, \bibinfo{person}{Xiaofei Wu}, \bibinfo{person}{Songcen Xu}, \bibinfo{person}{Youliang Yan}, {and} \bibinfo{person}{Wenming Yang}.} \bibinfo{year}{2024}\natexlab{a}.
\newblock \showarticletitle{VastGaussian: Vast 3D Gaussians for Large Scene Reconstruction}. In \bibinfo{booktitle}{\emph{Proceedings of the IEEE/CVF Conference on Computer Vision and Pattern Recognition (CVPR)}}. \bibinfo{pages}{5166--5175}.
\newblock


\bibitem[Lorensen and Cline(1998)]%
        {lorensen1998marching}
\bibfield{author}{\bibinfo{person}{William~E Lorensen} {and} \bibinfo{person}{Harvey~E Cline}.} \bibinfo{year}{1998}\natexlab{}.
\newblock \showarticletitle{Marching cubes: A high resolution 3D surface construction algorithm}.
\newblock In \bibinfo{booktitle}{\emph{Seminal graphics: pioneering efforts that shaped the field}}. \bibinfo{pages}{347--353}.
\newblock


\bibitem[Luo et~al\mbox{.}(2016)]%
        {Luo2016}
\bibfield{author}{\bibinfo{person}{Wenjie Luo}, \bibinfo{person}{Alexander~G. Schwing}, {and} \bibinfo{person}{Raquel Urtasun}.} \bibinfo{year}{2016}\natexlab{}.
\newblock \showarticletitle{Computing the Stereo Matching Cost with a Convolutional Neural Network}. In \bibinfo{booktitle}{\emph{Conference on Computer Vision and Pattern Recognition (CVPR)}}.
\newblock


\bibitem[Mildenhall et~al\mbox{.}(2021)]%
        {mildenhall2021nerf}
\bibfield{author}{\bibinfo{person}{Ben Mildenhall}, \bibinfo{person}{Pratul~P Srinivasan}, \bibinfo{person}{Matthew Tancik}, \bibinfo{person}{Jonathan~T Barron}, \bibinfo{person}{Ravi Ramamoorthi}, {and} \bibinfo{person}{Ren Ng}.} \bibinfo{year}{2021}\natexlab{}.
\newblock \showarticletitle{Nerf: Representing scenes as neural radiance fields for view synthesis}.
\newblock \bibinfo{journal}{\emph{Commun. ACM}} \bibinfo{volume}{65}, \bibinfo{number}{1} (\bibinfo{year}{2021}), \bibinfo{pages}{99--106}.
\newblock


\bibitem[M{\"u}ller et~al\mbox{.}(2022)]%
        {muller2022instant}
\bibfield{author}{\bibinfo{person}{Thomas M{\"u}ller}, \bibinfo{person}{Alex Evans}, \bibinfo{person}{Christoph Schied}, {and} \bibinfo{person}{Alexander Keller}.} \bibinfo{year}{2022}\natexlab{}.
\newblock \showarticletitle{Instant neural graphics primitives with a multiresolution hash encoding}.
\newblock \bibinfo{journal}{\emph{ACM transactions on graphics (TOG)}} \bibinfo{volume}{41}, \bibinfo{number}{4} (\bibinfo{year}{2022}), \bibinfo{pages}{1--15}.
\newblock


\bibitem[Newcombe et~al\mbox{.}(2011)]%
        {newcombe2011kinectfusion}
\bibfield{author}{\bibinfo{person}{Richard~A Newcombe}, \bibinfo{person}{Shahram Izadi}, \bibinfo{person}{Otmar Hilliges}, \bibinfo{person}{David Molyneaux}, \bibinfo{person}{David Kim}, \bibinfo{person}{Andrew~J Davison}, \bibinfo{person}{Pushmeet Kohi}, \bibinfo{person}{Jamie Shotton}, \bibinfo{person}{Steve Hodges}, {and} \bibinfo{person}{Andrew Fitzgibbon}.} \bibinfo{year}{2011}\natexlab{}.
\newblock \showarticletitle{Kinectfusion: Real-time dense surface mapping and tracking}. In \bibinfo{booktitle}{\emph{2011 10th IEEE international symposium on mixed and augmented reality}}. Ieee, \bibinfo{pages}{127--136}.
\newblock


\bibitem[Reiser et~al\mbox{.}(2023)]%
        {reiser2023merf}
\bibfield{author}{\bibinfo{person}{Christian Reiser}, \bibinfo{person}{Rick Szeliski}, \bibinfo{person}{Dor Verbin}, \bibinfo{person}{Pratul Srinivasan}, \bibinfo{person}{Ben Mildenhall}, \bibinfo{person}{Andreas Geiger}, \bibinfo{person}{Jon Barron}, {and} \bibinfo{person}{Peter Hedman}.} \bibinfo{year}{2023}\natexlab{}.
\newblock \showarticletitle{Merf: Memory-efficient radiance fields for real-time view synthesis in unbounded scenes}.
\newblock \bibinfo{journal}{\emph{ACM Transactions on Graphics (TOG)}} \bibinfo{volume}{42}, \bibinfo{number}{4} (\bibinfo{year}{2023}), \bibinfo{pages}{1--12}.
\newblock


\bibitem[Sch\"{o}nberger and Frahm(2016)]%
        {schoenberger2016sfm}
\bibfield{author}{\bibinfo{person}{Johannes~Lutz Sch\"{o}nberger} {and} \bibinfo{person}{Jan-Michael Frahm}.} \bibinfo{year}{2016}\natexlab{}.
\newblock \showarticletitle{Structure-from-Motion Revisited}. In \bibinfo{booktitle}{\emph{Conference on Computer Vision and Pattern Recognition (CVPR)}}.
\newblock


\bibitem[Sun et~al\mbox{.}(2003)]%
        {Sun2003}
\bibfield{author}{\bibinfo{person}{Jian Sun}, \bibinfo{person}{Nan-Ning Zheng}, {and} \bibinfo{person}{Heung-Yeung Shum}.} \bibinfo{year}{2003}\natexlab{}.
\newblock \showarticletitle{Stereo matching using belief propagation}.
\newblock \bibinfo{journal}{\emph{TPAMI}}  \bibinfo{volume}{25} (\bibinfo{year}{2003}), \bibinfo{pages}{787 -- 800}.
\newblock
Issue 7.


\bibitem[Vespa et~al\mbox{.}(2019)]%
        {vespa2019adaptive}
\bibfield{author}{\bibinfo{person}{Emanuele Vespa}, \bibinfo{person}{Nils Funk}, \bibinfo{person}{Paul~HJ Kelly}, {and} \bibinfo{person}{Stefan Leutenegger}.} \bibinfo{year}{2019}\natexlab{}.
\newblock \showarticletitle{Adaptive-resolution octree-based volumetric SLAM}. In \bibinfo{booktitle}{\emph{2019 International Conference on 3D Vision (3DV)}}. IEEE, \bibinfo{pages}{654--662}.
\newblock


\bibitem[Wang et~al\mbox{.}(2022)]%
        {wang2022neuris}
\bibfield{author}{\bibinfo{person}{Jiepeng Wang}, \bibinfo{person}{Peng Wang}, \bibinfo{person}{Xiaoxiao Long}, \bibinfo{person}{Christian Theobalt}, \bibinfo{person}{Taku Komura}, \bibinfo{person}{Lingjie Liu}, {and} \bibinfo{person}{Wenping Wang}.} \bibinfo{year}{2022}\natexlab{}.
\newblock \showarticletitle{Neuris: Neural reconstruction of indoor scenes using normal priors}. In \bibinfo{booktitle}{\emph{European Conference on Computer Vision}}. Springer, \bibinfo{pages}{139--155}.
\newblock


\bibitem[Wang et~al\mbox{.}(2024)]%
        {wang2024dc}
\bibfield{author}{\bibinfo{person}{Linhan Wang}, \bibinfo{person}{Kai Cheng}, \bibinfo{person}{Shuo Lei}, \bibinfo{person}{Shengkun Wang}, \bibinfo{person}{Wei Yin}, \bibinfo{person}{Chenyang Lei}, \bibinfo{person}{Xiaoxiao Long}, {and} \bibinfo{person}{Chang-Tien Lu}.} \bibinfo{year}{2024}\natexlab{}.
\newblock \showarticletitle{DC-Gaussian: Improving 3D Gaussian Splatting for Reflective Dash Cam Videos}.
\newblock \bibinfo{journal}{\emph{arXiv preprint arXiv:2405.17705}} (\bibinfo{year}{2024}).
\newblock


\bibitem[Wang et~al\mbox{.}(2021)]%
        {wang2021neus}
\bibfield{author}{\bibinfo{person}{Peng Wang}, \bibinfo{person}{Lingjie Liu}, \bibinfo{person}{Yuan Liu}, \bibinfo{person}{Christian Theobalt}, \bibinfo{person}{Taku Komura}, {and} \bibinfo{person}{Wenping Wang}.} \bibinfo{year}{2021}\natexlab{}.
\newblock \showarticletitle{Neus: Learning neural implicit surfaces by volume rendering for multi-view reconstruction}.
\newblock \bibinfo{journal}{\emph{arXiv preprint arXiv:2106.10689}} (\bibinfo{year}{2021}).
\newblock


\bibitem[Yao et~al\mbox{.}(2018)]%
        {yao2018mvsnet}
\bibfield{author}{\bibinfo{person}{Yao Yao}, \bibinfo{person}{Zixin Luo}, \bibinfo{person}{Shiwei Li}, \bibinfo{person}{Tian Fang}, {and} \bibinfo{person}{Long Quan}.} \bibinfo{year}{2018}\natexlab{}.
\newblock \showarticletitle{Mvsnet: Depth inference for unstructured multi-view stereo}. In \bibinfo{booktitle}{\emph{Proceedings of the European conference on computer vision (ECCV)}}. \bibinfo{pages}{767--783}.
\newblock


\bibitem[Yariv et~al\mbox{.}(2021)]%
        {yariv2021volume}
\bibfield{author}{\bibinfo{person}{Lior Yariv}, \bibinfo{person}{Jiatao Gu}, \bibinfo{person}{Yoni Kasten}, {and} \bibinfo{person}{Yaron Lipman}.} \bibinfo{year}{2021}\natexlab{}.
\newblock \showarticletitle{Volume rendering of neural implicit surfaces}.
\newblock \bibinfo{journal}{\emph{Advances in Neural Information Processing Systems}}  \bibinfo{volume}{34} (\bibinfo{year}{2021}), \bibinfo{pages}{4805--4815}.
\newblock


\bibitem[Yu et~al\mbox{.}(2021)]%
        {yu2021plenoctrees}
\bibfield{author}{\bibinfo{person}{Alex Yu}, \bibinfo{person}{Ruilong Li}, \bibinfo{person}{Matthew Tancik}, \bibinfo{person}{Hao Li}, \bibinfo{person}{Ren Ng}, {and} \bibinfo{person}{Angjoo Kanazawa}.} \bibinfo{year}{2021}\natexlab{}.
\newblock \showarticletitle{Plenoctrees for real-time rendering of neural radiance fields}. In \bibinfo{booktitle}{\emph{Proceedings of the IEEE/CVF International Conference on Computer Vision}}. \bibinfo{pages}{5752--5761}.
\newblock


\bibitem[Yu et~al\mbox{.}(2024b)]%
        {yu2024gsdf}
\bibfield{author}{\bibinfo{person}{Mulin Yu}, \bibinfo{person}{Tao Lu}, \bibinfo{person}{Linning Xu}, \bibinfo{person}{Lihan Jiang}, \bibinfo{person}{Yuanbo Xiangli}, {and} \bibinfo{person}{Bo Dai}.} \bibinfo{year}{2024}\natexlab{b}.
\newblock \showarticletitle{GSDF: 3DGS Meets SDF for Improved Rendering and Reconstruction}.
\newblock \bibinfo{journal}{\emph{arXiv preprint arXiv:2403.16964}} (\bibinfo{year}{2024}).
\newblock


\bibitem[Yu et~al\mbox{.}(2024a)]%
        {yu2023mip}
\bibfield{author}{\bibinfo{person}{Zehao Yu}, \bibinfo{person}{Anpei Chen}, \bibinfo{person}{Binbin Huang}, \bibinfo{person}{Torsten Sattler}, {and} \bibinfo{person}{Andreas Geiger}.} \bibinfo{year}{2024}\natexlab{a}.
\newblock \showarticletitle{Mip-Splatting: Alias-free 3D Gaussian Splatting}.
\newblock \bibinfo{journal}{\emph{Conference on Computer Vision and Pattern Recognition (CVPR)}} (\bibinfo{year}{2024}).
\newblock


\bibitem[Yu et~al\mbox{.}(2022)]%
        {yu2022monosdf}
\bibfield{author}{\bibinfo{person}{Zehao Yu}, \bibinfo{person}{Songyou Peng}, \bibinfo{person}{Michael Niemeyer}, \bibinfo{person}{Torsten Sattler}, {and} \bibinfo{person}{Andreas Geiger}.} \bibinfo{year}{2022}\natexlab{}.
\newblock \showarticletitle{Monosdf: Exploring monocular geometric cues for neural implicit surface reconstruction}.
\newblock \bibinfo{journal}{\emph{Advances in neural information processing systems}}  \bibinfo{volume}{35} (\bibinfo{year}{2022}), \bibinfo{pages}{25018--25032}.
\newblock


\bibitem[Yu et~al\mbox{.}(2024c)]%
        {yu2024gaussian}
\bibfield{author}{\bibinfo{person}{Zehao Yu}, \bibinfo{person}{Torsten Sattler}, {and} \bibinfo{person}{Andreas Geiger}.} \bibinfo{year}{2024}\natexlab{c}.
\newblock \showarticletitle{Gaussian Opacity Fields: Efficient and Compact Surface Reconstruction in Unbounded Scenes}.
\newblock \bibinfo{journal}{\emph{arXiv preprint arXiv:2404.10772}} (\bibinfo{year}{2024}).
\newblock


\bibitem[Zach et~al\mbox{.}(2011)]%
        {Zach2007}
\bibfield{author}{\bibinfo{person}{Christopher Zach}, \bibinfo{person}{Thomas Pock}, {and} \bibinfo{person}{Horst Bischof}.} \bibinfo{year}{2011}\natexlab{}.
\newblock \showarticletitle{A Globally Optimal Algorithm for Robust TV-L1 Range Image Integration}. In \bibinfo{booktitle}{\emph{International Conference on Computer Vision (ICCV)}}.
\newblock


\bibitem[Zbontar and LeCun(2015)]%
        {Zbontar2015}
\bibfield{author}{\bibinfo{person}{Jure Zbontar} {and} \bibinfo{person}{Yann LeCun}.} \bibinfo{year}{2015}\natexlab{}.
\newblock \showarticletitle{Efficient Deep Learning for Stereo Matching}. In \bibinfo{booktitle}{\emph{Conference on Computer Vision and Pattern Recognition (CVPR)}}.
\newblock


\bibitem[Zwicker et~al\mbox{.}(2002)]%
        {zwicker2002ewa}
\bibfield{author}{\bibinfo{person}{Matthias Zwicker}, \bibinfo{person}{Hanspeter Pfister}, \bibinfo{person}{Jeroen Van~Baar}, {and} \bibinfo{person}{Markus Gross}.} \bibinfo{year}{2002}\natexlab{}.
\newblock \showarticletitle{EWA splatting}.
\newblock \bibinfo{journal}{\emph{IEEE Transactions on Visualization and Computer Graphics}} \bibinfo{volume}{8}, \bibinfo{number}{3} (\bibinfo{year}{2002}), \bibinfo{pages}{223--238}.
\newblock


\end{thebibliography}

\end{document}


\title{Appendix: RaDe-GS: Rasterizing Depth in Gaussian Splatting}


\maketitle

\section{Details on Ray Space Depth}
In this section, we will show more derivation of depth rasterization.
In the main submission, we formulate Gaussian depth as:
\begin{equation}
    d = z_c + \mathbf{p} \begin{pmatrix}\Delta u\\
                                                \Delta v
                                                \end{pmatrix},
    \label{eq:depth}
\end{equation}
where $z_c$ is the depth of the Gaussian center, $\Delta u = u_c - u$ and $\Delta v = v_c - v$ are the relative pixel positions.
We have derived the second term $\mathbf{p} \begin{pmatrix}\Delta u\\\Delta v\end{pmatrix}$, and now we show how we reach the first term $z_c$.

In Equation~16 of the main submission, we have divided the depth of a Gaussian into two parts:
\begin{equation}
    d = \hat{\mathbf{p}} \left(\begin{matrix}0 \\0 \\t_c\end{matrix}\right) + \hat{\mathbf{p}} \left(\begin{matrix}\Delta u \\\Delta v \\0\end{matrix}\right),
    \label{eq:main_sub_depth}
\end{equation}
where $\hat{\mathbf{p}}$ is in form of:
\begin{equation}
    \hat{\mathbf{p}} = \frac{z_c}{t_c}\frac{\mathbf{v}'^\top \mathbf{\Sigma}'^{-1}}{\mathbf{v}'^\top \mathbf{\Sigma}'^{-1} \mathbf{v}'},
\label{eq:p_hat}
\end{equation}
where $t_c$ is the distance from camera center to Gaussian center, $\mathbf{\Sigma}'$ is the Gaussian covariance in ray space, and $\mathbf{v}'=(0,0,1)^\top$ is a constant vector in ray space.

We will simplify the $\hat{\mathbf{p}}(0,0,t_c)^T$ of Equation~\ref{eq:main_sub_depth}.
By substituting Equation~\ref{eq:p_hat} into the first part of Equation~\ref{eq:main_sub_depth}, we get:
\begin{equation}
\begin{split}
    \hat{\mathbf{p}} \left(\begin{matrix}0 \\0 \\t_c\end{matrix}\right)&=\frac{z_c}{t_c}\frac{\mathbf{v}'^\top \mathbf{\Sigma}'^{-1}}{\mathbf{v}'^\top \mathbf{\Sigma}'^{-1} \mathbf{v}'}\left(\begin{matrix}0 \\0 \\t_c\end{matrix}\right)\\
    &=\frac{z_c}{t_c}\ \frac{\mathbf{v}'^\top \mathbf{\Sigma}'^{-1}}{\mathbf{v}'^\top \mathbf{\Sigma}'^{-1} \mathbf{v}'}(t_c\mathbf{v}')\\
    &=\frac{z_c}{t_c}\ \frac{\mathbf{v}'^\top \mathbf{\Sigma}'^{-1}\mathbf{v}'}{\mathbf{v}'^\top \mathbf{\Sigma}'^{-1} \mathbf{v}'}t_c\\ &= z_c.
\end{split}
\end{equation}

\section{Details on Gaussian Normal}
In this section, we show the detailed derivation of Equation~18 in the main submission. Given image plane coordinates $(u,v)$, ray from $(u,v)$ intersects with Gaussian on the point $\mathbf{u}$.
\begin{equation}
    \mathbf{u} = 
    \left(\begin{array}{c}
    u \\
    v \\
    t^*
    \end{array}\right).
\end{equation}
As shown in the main submission, $t^*=\frac{t_c}{z_c}d$, and $d$ is shown in Equation~\ref{eq:depth}.
We plug in and get that:

\begin{equation}
\begin{split}
    \mathbf{u} &= 
    \left(\begin{array}{c}
    u \\
    v \\
    t^*
    \end{array}\right)=\left(\begin{array}{c}
    u \\
    v \\
    \frac{t_c}{z_c}d
    \end{array}\right)\\&=\left(\begin{array}{c}
    u_c - \Delta u \\
    v_c - \Delta v \\
    t_c +  (\Delta u \quad \Delta v)     \mathbf{q}^\top
    \end{array}\right)
    = \left(\begin{array}{c}
    u_c \\
    v_c \\
    t_c
    \end{array}\right) + 
    \left(\begin{array}{c}
    -\Delta u \\
    -\Delta v \\
     (\Delta u \quad \Delta v)     \mathbf{q}^\top
    \end{array}\right),
\end{split}
\label{eq:normal1}
\end{equation}
where $\mathbf{q}=\frac{t_c}{z_c}\mathbf{p}$. 
Since $\hat{\mathbf{q}}=\frac{t_c}{z_c}\hat{\mathbf{p}}$ (Equation~15 in the main submission) and $\mathbf{p}$ contains the first two components of $\hat{\mathbf{p}}$, we know that $\mathbf{q}$ is the same as $\hat{\mathbf{q}}$ skipping the last component.
